\newif\ifemulateapj
\emulateapjtrue        % use emulateapj format

\ifemulateapj
    \documentclass[iop]{emulateapj}
\else
    \documentclass[12pt,preprint]{aastex}
\fi
\newcommand{\tnm}[1]{\tablenotemark{#1}}
\newcommand{\peras}{\ensuremath{\mathrm{arcsec}^{-1}}}
\newcommand{\kms}{\mbox{km\ s$^{-1}$}}
\newcommand{\kmsMpc}{\kms~\mbox{Mpc}$^{-1}$}
\newcommand{\Msun}{\ensuremath{M_{\odot}}}
\newcommand{\Msunyr}{\Msun~\mbox{yr}$^{-1}$}
\newcommand{\Lstar}{\ensuremath{L_{*}}}
\newcommand{\Mstar}{\ensuremath{M_{*}}}
\newcommand{\multasec}[2]{\mbox{#1\arcsec$\,\times\,$#2\arcsec}}

\newcommand{\dg}{\ensuremath{^\circ}}
\newcommand{\HST}{\textit{HST}}
\newcommand{\Spitzer}{\textit{Spitzer}}
\newcommand{\Vband}{\ensuremath{V_{606}}}
\newcommand{\Vbandb}{\ensuremath{V_{600LP}}}
\newcommand{\Yband}{\ensuremath{Y_{098}}}
\newcommand{\Jband}{\ensuremath{J_{125}}}
\newcommand{\Hband}{\ensuremath{H_{160}}}

\newcommand{\YmJ}{\ensuremath{\Yband - \Jband}}
\newcommand{\JmH}{\ensuremath{\Jband - \Hband}}
\newcommand{\ionrm}[1]{\mbox{\small\sc{\romannumeral #1}}}
\newcommand{\forb}[2]{\mbox{[#1~\ionrm{#2}]}}
\newcommand{\CII}{\forb{C}{2}}

\def\figa {
    \begin{figure}[t]
    \ifemulateapj
        \epsscale{1.2}
        \plotone{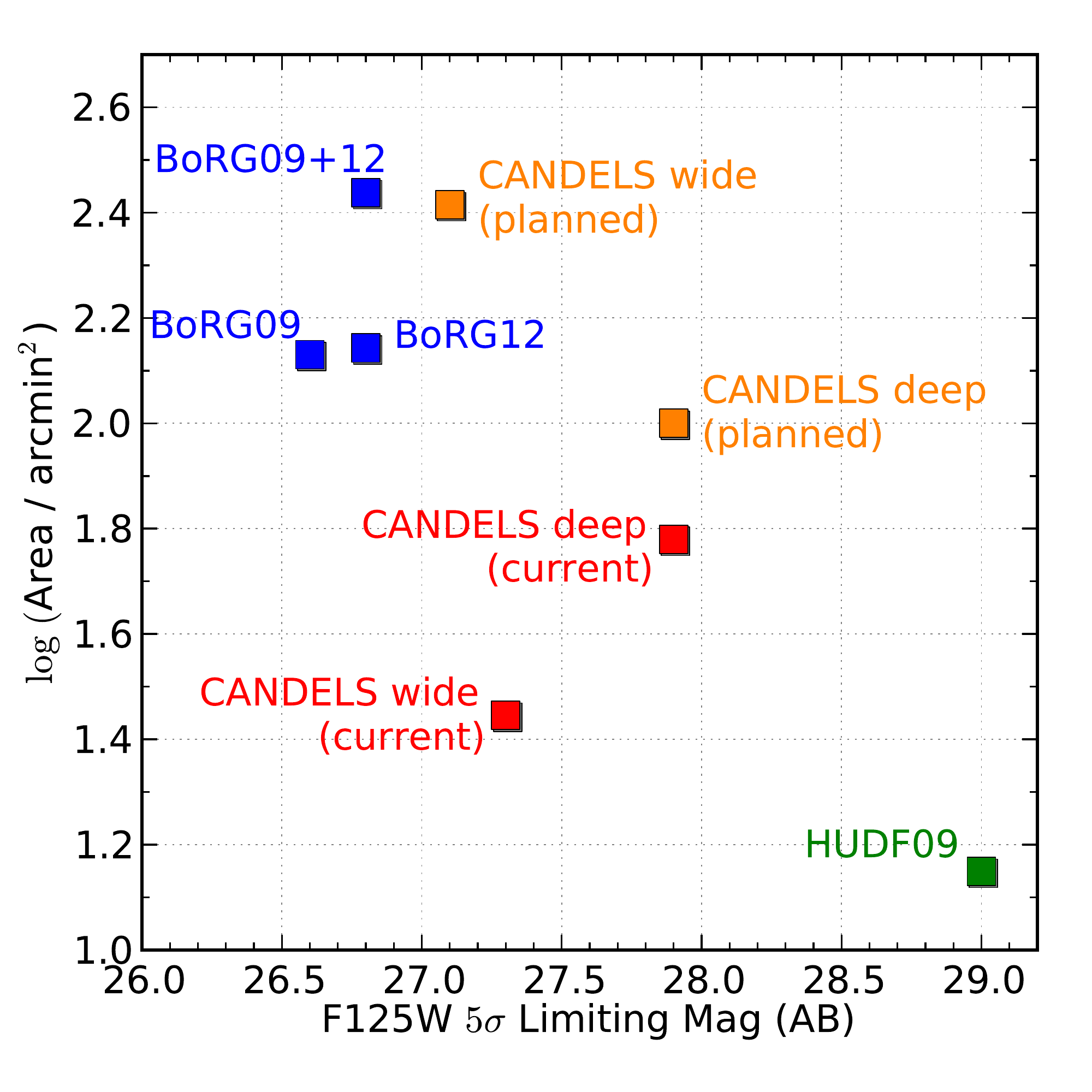}
    \else
        \epsscale{1.0}
        \plotone{f1}
    \fi
    \caption{Comparison of area and depth of current surveys that can search for $z\sim8$ LBGs.  Because only $\sim260$~arcmin$^{2}$ of the planned CANDELS wide survey includes $Y$-band imaging (red and orange), the ongoing BoRG survey (blue; $\sim274$~arcmin$^{2}$) provides the largest area for a $Y$-band dropout survey.  The depth of our BoRG survey is only $\sim0.3$ mag shallower than the wide part of the CANDELS survey.  The ultradeep HUDF09 WFC3/IR observations are illustrated by the green point.}
    \label{fig:survey}
    \end{figure}
}

\def\figb {
    \begin{figure*}[ht]
    \ifemulateapj
        \epsscale{0.75}
        \plotone{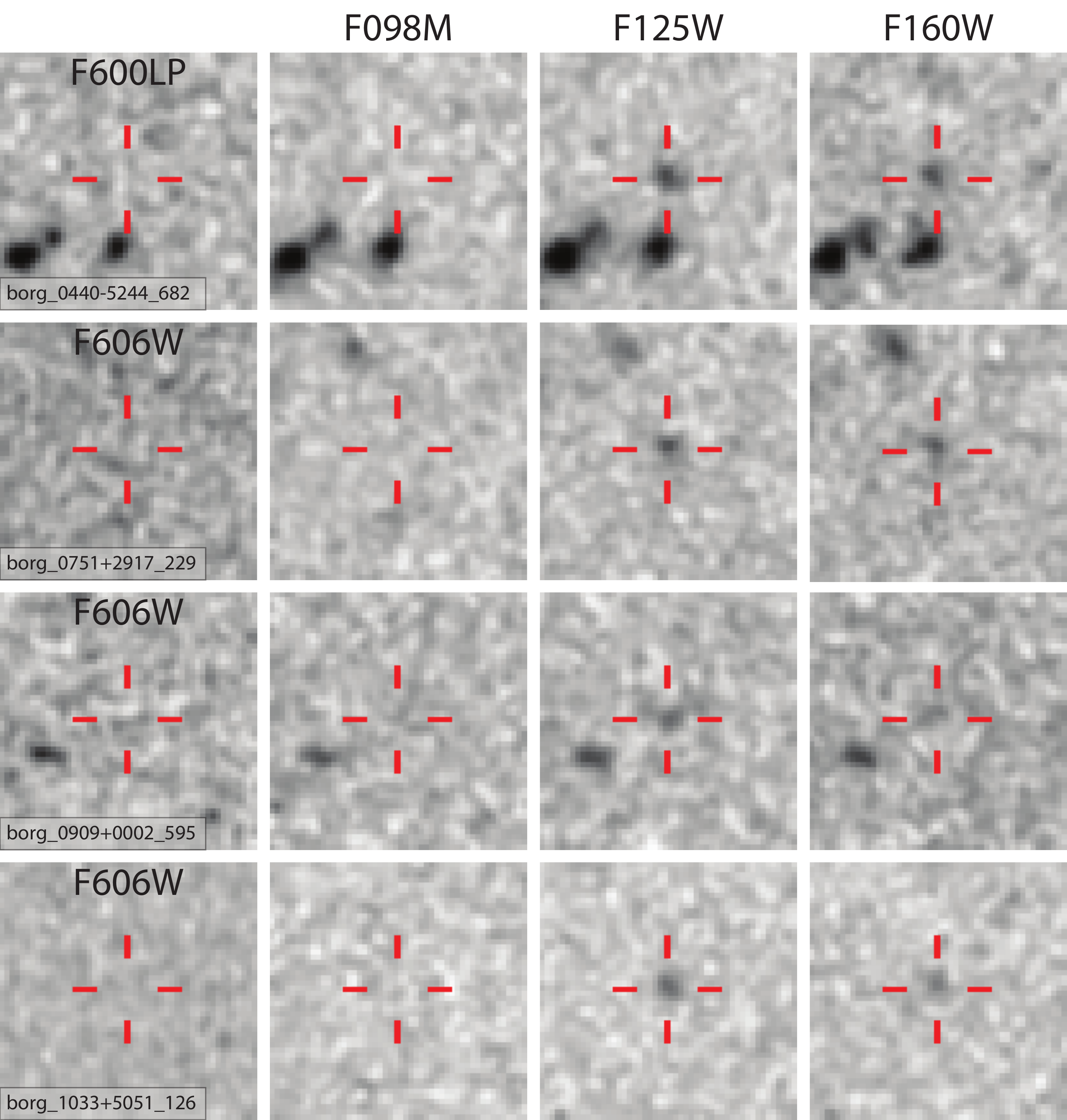}
    \else
        \epsscale{1.0}
        \plotone{f2}
    \fi
    \caption{Postage-stamp cutout images of four of the very bright ($\ga8\sigma$) high-redshift \Yband-dropout candidate galaxies.  The cutout images are \multasec{4}{4}, corresponding to 19.2~kpc on a side at $z=8$, and are shown with a P.A.$=0\dg$.}
    \label{fig:stamps1}
    \end{figure*}
}

\def\figc {
    \begin{figure*}[ht]
    \ifemulateapj
        \epsscale{0.75}
        \plotone{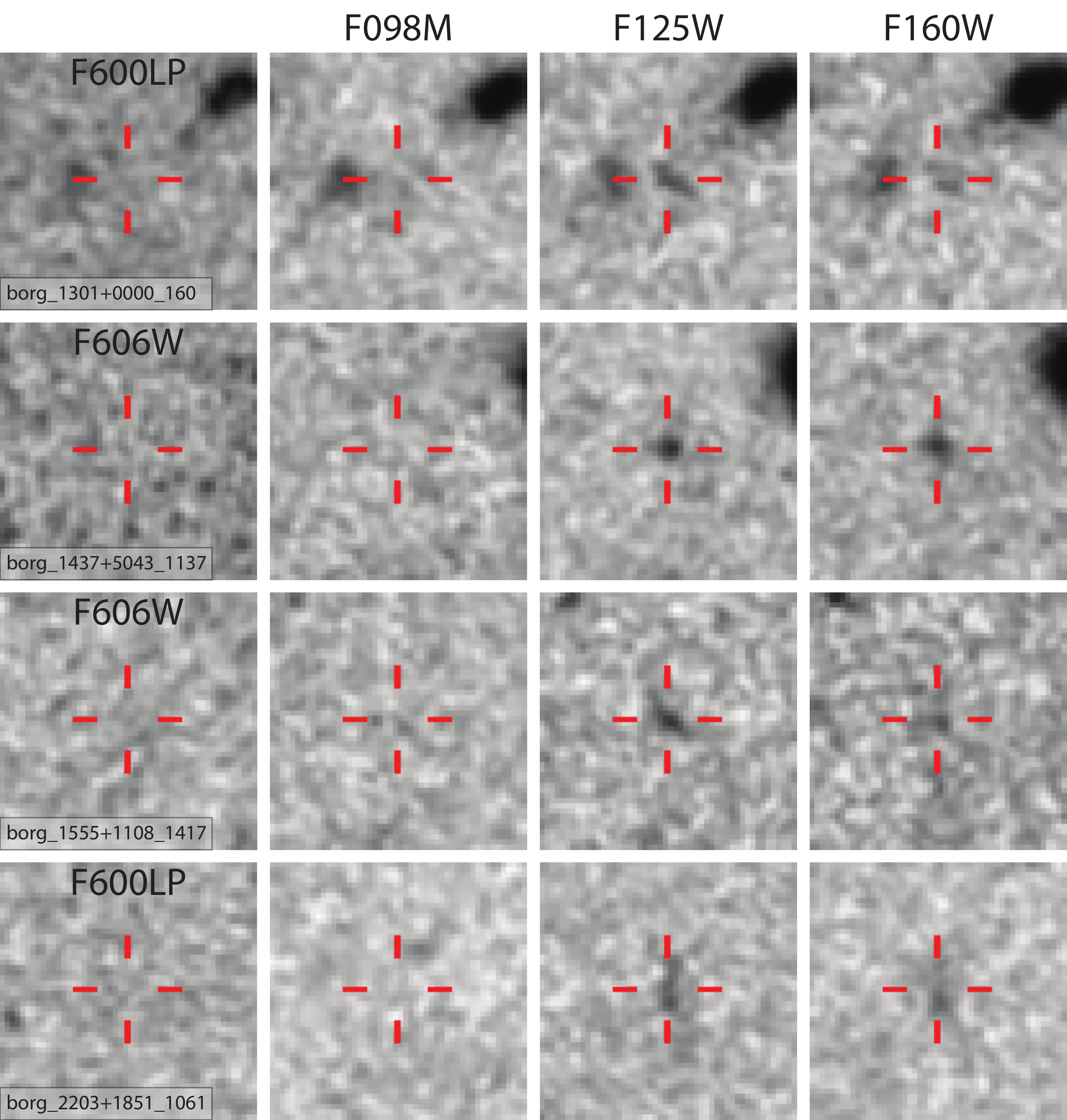}
    \else
        \epsscale{1.0}
        \plotone{f3}
    \fi
    \caption{Postage-stamp cutout images of four of the very bright ($\ga8\sigma$) high-redshift \Yband-dropout candidate galaxies.  The cutout images are \multasec{4}{4}, corresponding to 19.2~kpc on a side at $z=8$, and are shown with a P.A.$=0\dg$.}
    \label{fig:stamps2}
    \end{figure*}
}

\def\figd {
    \begin{figure}[ht]
    \ifemulateapj
        \epsscale{1.2}
        \plotone{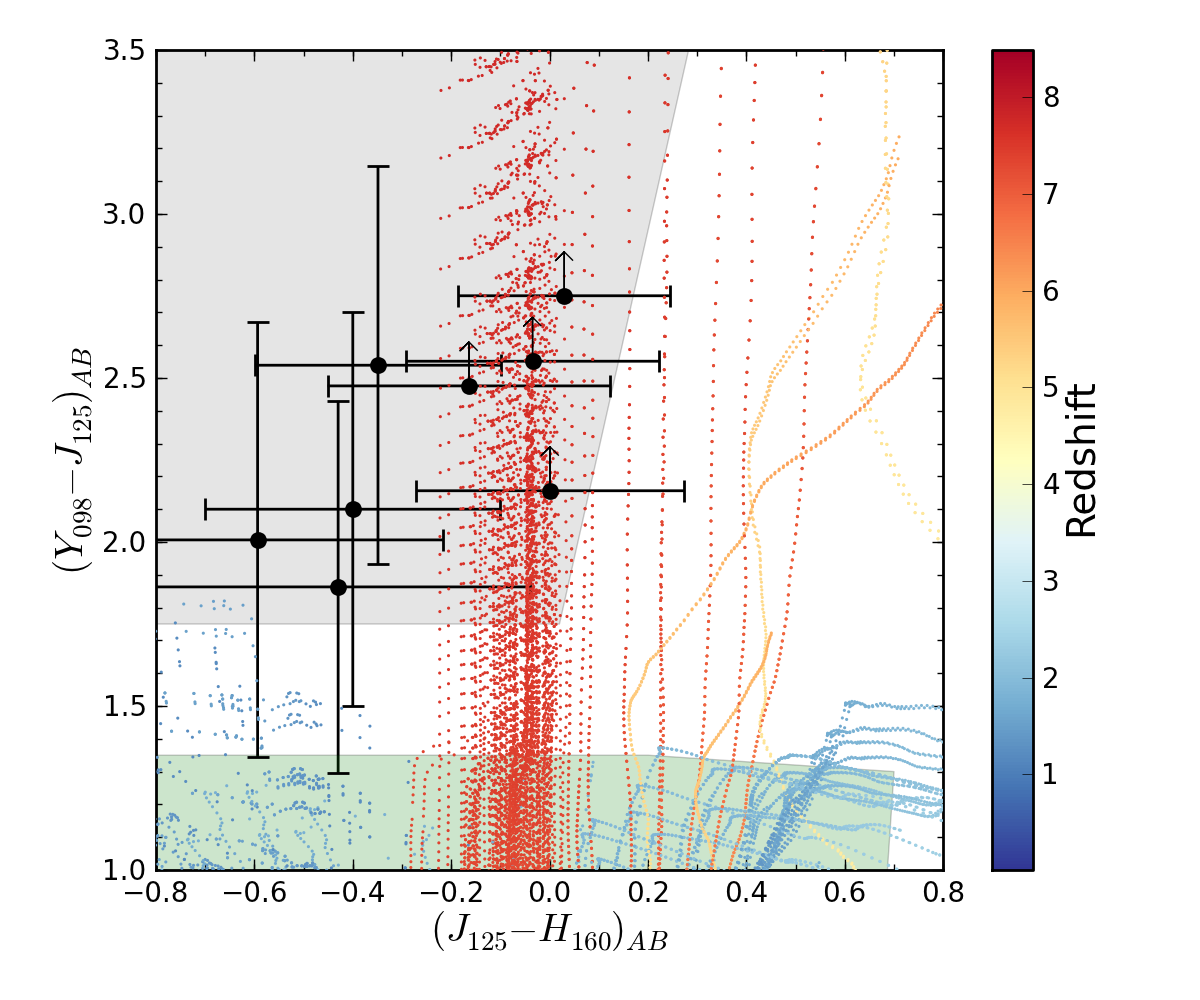}
    \else
        \epsscale{1.0}
        \plotone{f4}
    \fi
    \caption{\YmJ\ vs. \JmH\ two-color diagram used to select our \Yband-band dropout candidates.  The colors of our $8\sigma$ sources are shown by the black data points.  The error bars and lower limits are $1\sigma$ (68\% confidence).  The gray region represents the \YmJ\ and \JmH\ colors of our selection criteria.  The colored points represent the expected colors of galaxies simulated over a wide range in redshifts.  The green region indicates the colors of low-mass L,T dwarf stars (e.g., \citealt{Knapp2004, Ryan2011}; Holwerda et al. 2012 (in prep)).  In addition to the color-color selection shown here, \Yband-dropouts also need to be undetected ($S/N<1.5$) in the deep optical imaging performed in the $V$ band by the BoRG survey.  The non-detection in $V$ band is fundamental to produce a clean sample of $z\sim8$ candidates (see \citealt{Bouwens2011b}).}
    \label{fig:colcol}
    \end{figure}
}

\def\fige {
    \begin{figure}[t]
    \ifemulateapj
        \epsscale{1.2}
        \plotone{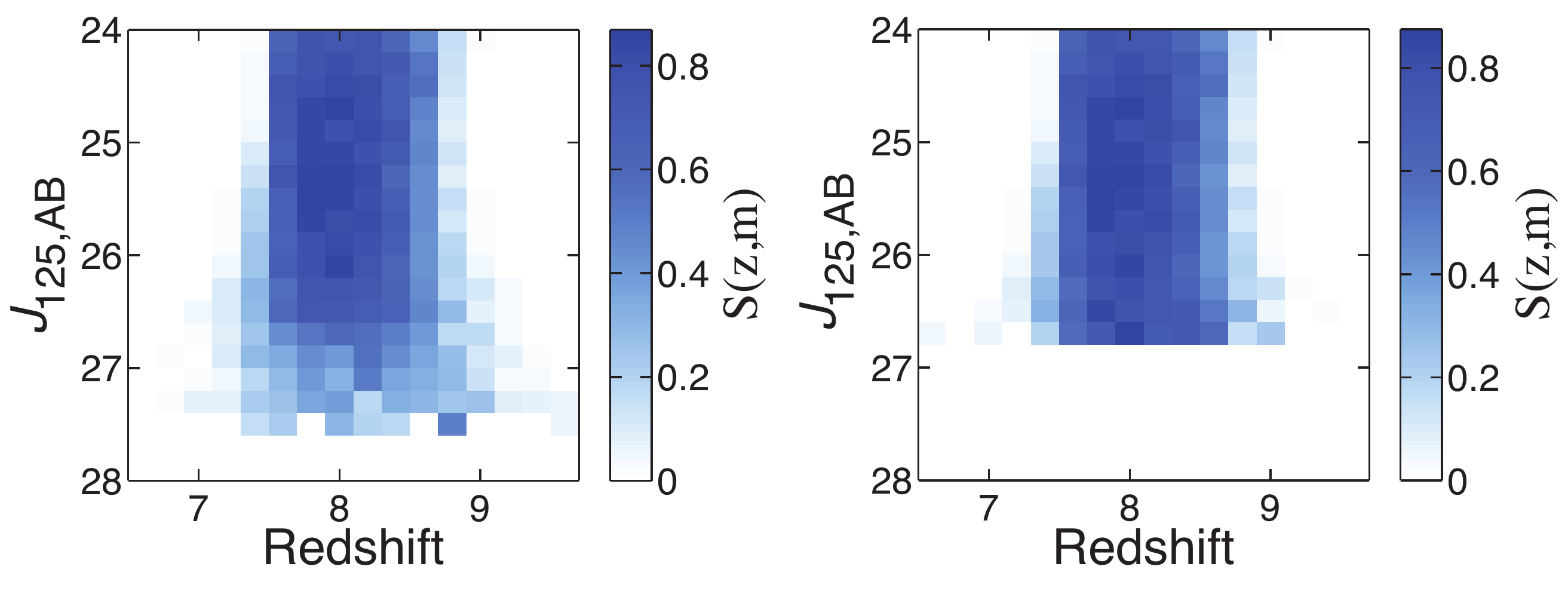}
    \else
        \epsscale{1.0}
        \plotone{f5}
    \fi
    \caption{The $S(z,m)$ magnitude-dependent redshift selection function of one representative BoRG survey field, borg\_$1437+5043$, for both the $5\sigma$ (left) and $8\sigma$ (right) $z\sim8$ source catalogs.  These selection functions were obtained through simulations to recover artificial sources in the BoRG images, as discussed in Section~\ref{sect:uvlf}.  Through these simulations, we computed $S(z,m)$ for each of the 59 individual BoRG fields.}
    \label{fig:szm}
    \end{figure}
}

\def\figf {
    \begin{figure}[tb]
    \ifemulateapj
        \epsscale{1.2}
        \plotone{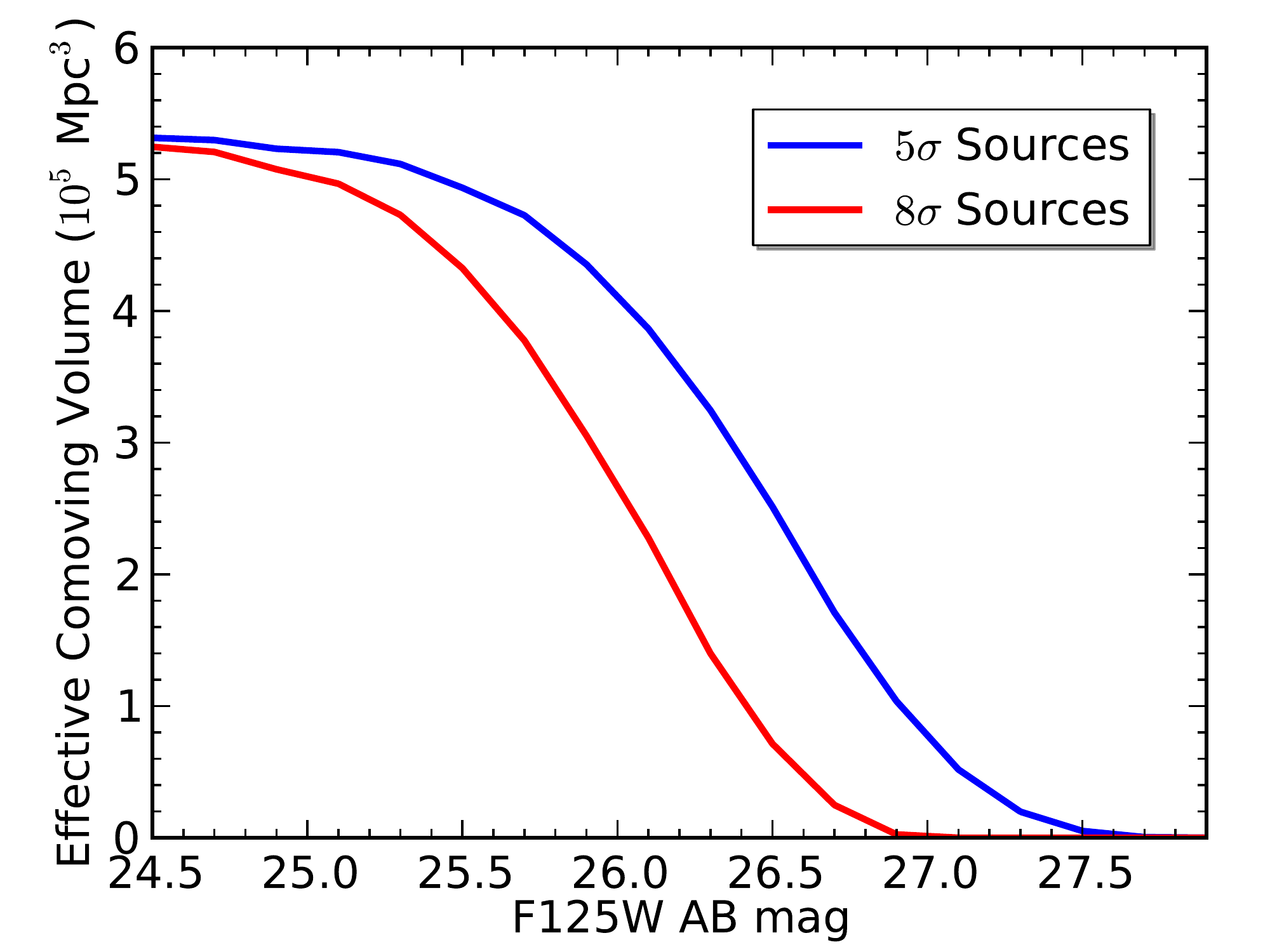}
    \else
        \epsscale{1.0}
        \plotone{f6}
    \fi
    \caption{Effective comoving volume for our $z\sim8$ selection as a function of \Jband\ magnitude for both the $5\sigma$ (blue) and $8\sigma$ (red) source catalogs.  The effective volume shown here takes into account reductions from both photometric scatter and incompleteness.}
    \label{fig:effvol}
    \end{figure}
}

\def\figg {
    \begin{figure}[t]
    \ifemulateapj
        \epsscale{1.2}
        \plotone{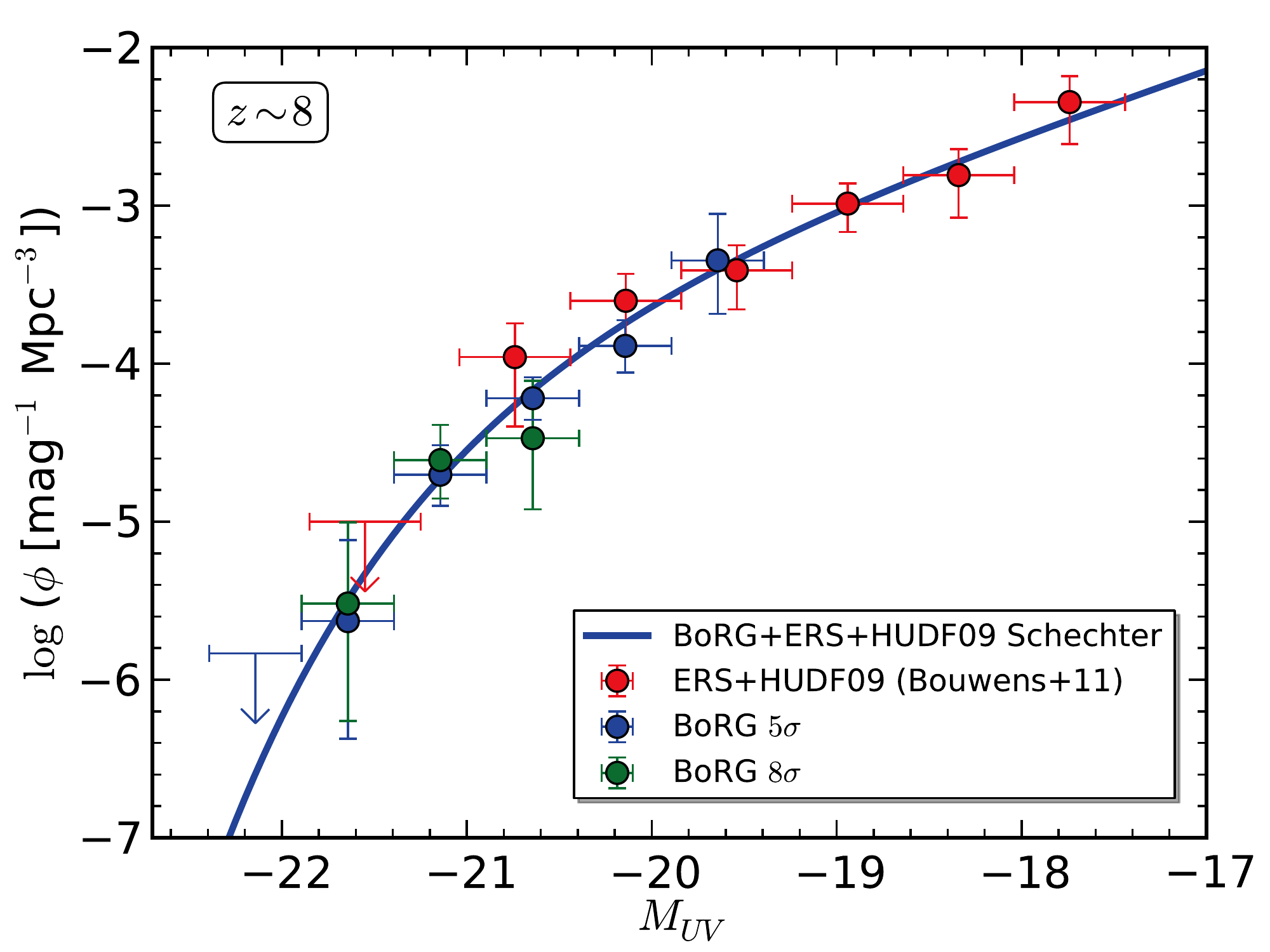}
    \else
        \epsscale{1.0}
        \plotone{f7}
    \fi
    \caption{The $z\sim8$ galaxy UV LF from the current BoRG dataset.  The stepwise LFs for our $5\sigma$ and $8\sigma$ catalogs are shown in blue and green, respectively.  The red points represent the stepwise LF derived by \cite{Bouwens2011b} for the ERS+HUDF09 dataset.  The blue line is the best-fit Schechter LF from combining the BoRG+ERS+HUDF09 dataset, providing the widest dynamic range in luminosity that is currently available.  Our results are consistent with a Schechter form of the UV LF and do not indicate an excess of bright $z\sim8$ LBGs.}
    \label{fig:uvlf}
    \end{figure}
}

\def\figh {
    \begin{figure}[ht]
    \ifemulateapj
        \epsscale{1.2}
        \plotone{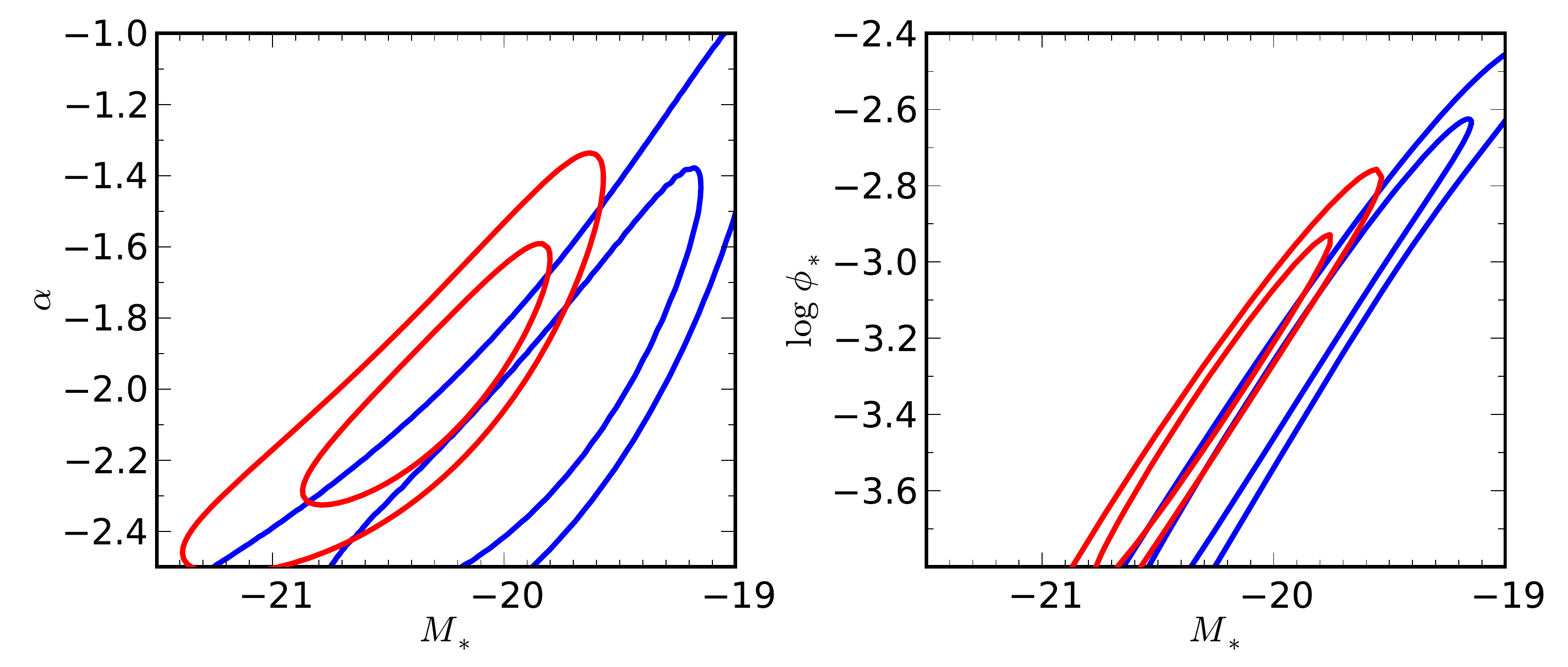}
    \else
        \epsscale{1.0}
        \plotone{f8}
    \fi
    \caption{$1\sigma\ (68\%)$ and $2\sigma\ (95\%)$ confidence intervals in the Schechter LF fit from BoRG+ERS+HUDF data (red lines) compared to the CANDELS+ERS+HUDF determination by \cite{Oesch2012} (blue lines).  The BoRG fit has a preference for a marginally brighter $M_{*}$, but the two datasets are consistent within their $2\sigma$ contours.  In particular, we note that our \Mstar\ versus $\alpha$ parameter values have much better constraints than the previous studies at $z\sim8$.}
    \label{fig:lfparam}
    \end{figure}
}

\def\figi {
    \begin{figure}[ht]
    \ifemulateapj
        \epsscale{1.2}
        \plotone{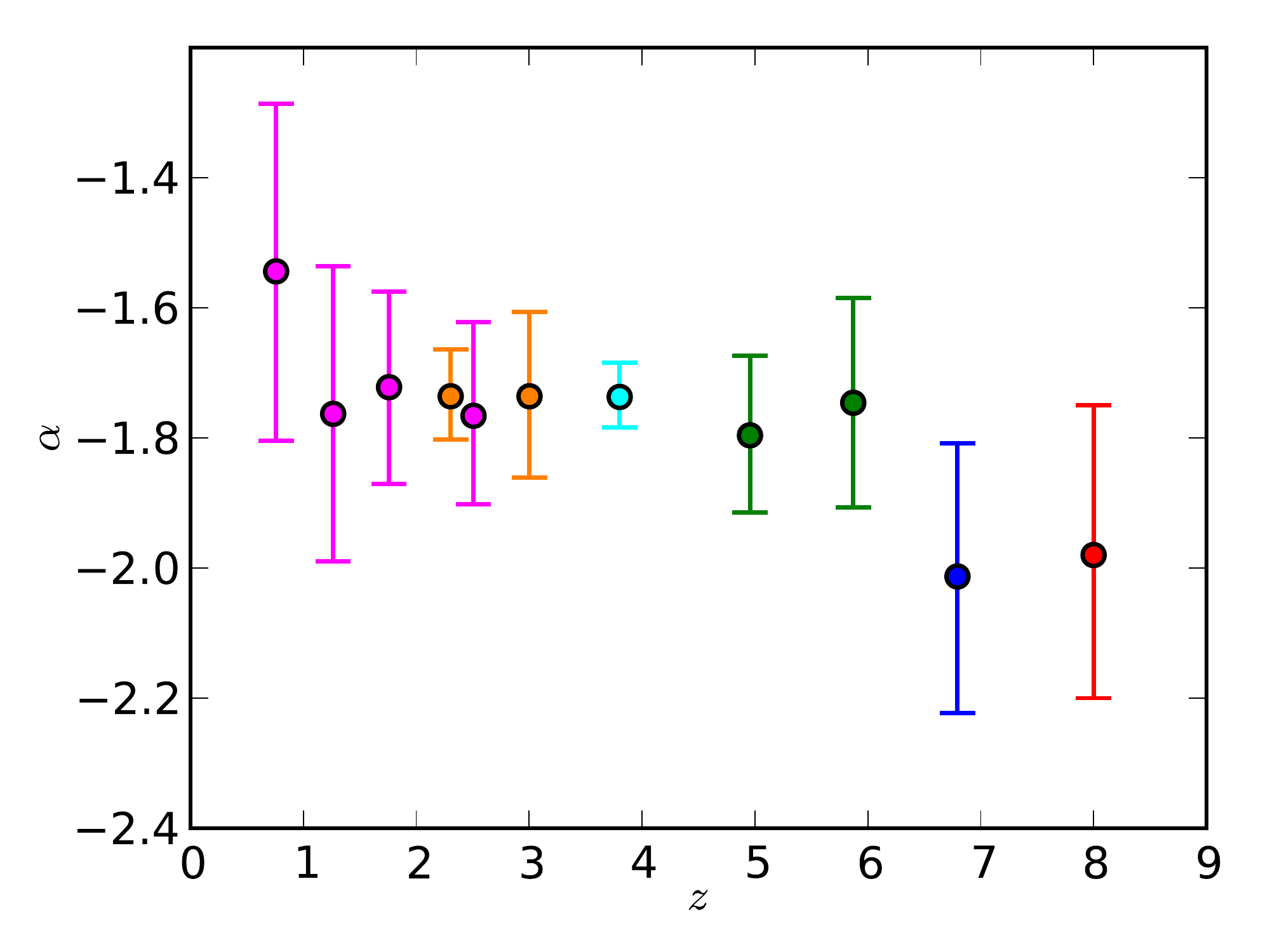}
    \else
        \epsscale{1.0}
        \plotone{f9}
    \fi
    \caption{Determination of the Schechter function faint-end slope $\alpha$ for the galaxy LF as a function of redshift. Literature determinations are from \cite{Oesch2010} at $z\sim0.7-2.5$ (magenta), from \cite{Reddy2009} at $z\sim2-3$ (orange), from \cite{Bouwens2007b} at $z\sim4$ (cyan), from \cite{Bouwens2012} at $z\sim5-6$ (green), and from \cite{Bouwens2011b} at $z\sim7$ (blue).  Our new determination at $z\sim8$ is shown in red.  All error bars are $1\sigma$ (68\% confidence).  Our latest determination provides some evidence that the LF is becoming steeper at $z\gtrsim7$, consistent with the $z\sim7$ measurement.}
    \label{fig:alpha}
    \end{figure}
}

\def\taba {
\ifemulateapj
    \begin{deluxetable*}{lclrrrrrrrrrrrr}
\else
    \begin{deluxetable}{lclrrrrrrrrrrrr}
    \rotate
    \tabletypesize{\scriptsize}
\fi
\tablecolumns{15}
\tablewidth{0pt}
\tablecaption{BoRG09 Survey Fields, Exposure Times, and $5\sigma$ Limiting Magnitudes\tnm{a}}
\tablehead{ \colhead{Field} & \colhead{Alternate} & \colhead{PID} & \colhead{$\alpha_{J2000}$}  & \colhead{$\delta_{J2000}$} & \multicolumn{2}{c}{F600LP} & \multicolumn{2}{c}{F606W} & \multicolumn{2}{c}{F098M} & \multicolumn{2}{c}{F125W} & \multicolumn{2}{c}{F160W} \\
\colhead{} & \colhead{Name\tnm{b}} & \colhead{} & \colhead{(deg)} & \colhead{(deg)} & \colhead{$t$ (s)} & \colhead{$m_{lim}$} & \colhead{$t$ (s)} & \colhead{$m_{lim}$} & \colhead{$t$ (s)} & \colhead{$m_{lim}$} & \colhead{$t$ (s)} & \colhead{$m_{lim}$} & \colhead{$t$ (s)} & \colhead{$m_{lim}$} }
\startdata
borg\_0214+1255 &  yan24 & 11702 & $ 33.410$ & $ 12.915$ & $ 2294$ & $25.9$ &         &        & $ 2806$ & $25.9$ & $ 1403$ & $25.9$ & $ 1403$ & $25.7$ \\
borg\_0540$-$6409 & borg2n & 11700 & $ 84.879$ & $-64.151$ &         &        & $ 3171$ & $26.9$ & $ 4112$ & $26.6$ & $ 2309$ & $26.6$ & $ 1406$ & $26.3$ \\
borg\_0553$-$6405 & borg81 & 11700 & $ 88.276$ & $-64.088$ &         &        & $ 3624$ & $27.0$ & $ 6418$ & $27.0$ & $ 2612$ & $26.8$ & $ 2012$ & $26.3$ \\
borg\_0624$-$6432 & borg2t\tnm{d} & 11700 & $ 95.898$ & $-64.528$ &         &        & $ 2133$ & $26.7$ & $ 1806$ & $26.3$ & $ 1206$ & $26.4$ & $  503$ & $25.6$ \\
borg\_0624$-$6440 & borg2k & 11700 & $ 95.951$ & $-64.663$ &         &        & $ 2135$ & $26.7$ & $ 2909$ & $26.6$ & $ 1206$ & $26.6$ & $  906$ & $26.0$ \\
borg\_0637$-$7518\tnm{c} & borg93 & 11700 & $ 99.265$ & $-75.313$ &         &        & $ 4290$ & $26.7$ & $ 6218$ & $26.7$ & $ 2412$ & $26.6$ & $ 1612$ & $26.0$ \\
borg\_0751+2917 & borg0t & 11700\tnm{e} & $117.709$ & $ 29.282$ & $ 3732$ & $26.6$ & $ 2826$ & $26.8$ & $18641$ & $27.4$ & $ 5115$ & $27.1$ & $ 3912$ & $26.8$ \\
borg\_0756+3043 & borg0c & 11700 & $118.989$ & $ 30.718$ &         &        & $ 2600$ & $26.7$ & $ 4712$ & $26.7$ & $ 1906$ & $26.6$ & $ 1406$ & $26.2$ \\
borg\_0808+3946 & borg1n & 11700 & $122.089$ & $ 39.759$ &         &        & $ 2600$ & $26.5$ & $ 4612$ & $26.4$ & $ 2206$ & $26.6$ & $ 1406$ & $26.0$ \\
borg\_0819+4911 & borg0g & 11700 & $124.830$ & $ 49.184$ &         &        & $ 1908$ & $26.4$ & $ 3009$ & $26.6$ & $ 1206$ & $26.5$ & $  806$ & $25.8$ \\
borg\_0820+2332 & borg30\tnm{d} & 11700 & $125.014$ & $ 23.535$ &         &        & $ 2556$ & $26.7$ & $ 3109$ & $26.5$ & $  703$ & $26.1$ & $  703$ & $25.8$ \\
borg\_0906+0255 & borg73 & 11700 & $136.405$ & $  2.925$ &         &        & $ 3106$ & $26.9$ & $ 5518$ & $27.0$ & $ 2709$ & $27.0$ & $ 1906$ & $26.6$ \\
borg\_0909+0002 & borg66 & 11700 & $137.286$ & $ -0.030$ &         &        & $ 2650$ & $26.8$ & $ 3909$ & $26.8$ & $ 1806$ & $26.7$ & $ 1006$ & $26.0$ \\
borg\_0914+2822 & borg39 & 11700 & $138.569$ & $ 28.362$ &         &        & $ 2571$ & $26.8$ & $ 4615$ & $26.8$ & $ 2206$ & $26.8$ & $ 1706$ & $26.5$ \\
borg\_0922+4505 & borg1r & 11700 & $140.406$ & $ 45.088$ &         &        & $ 2708$ & $26.6$ & $ 4812$ & $26.6$ & $ 2106$ & $26.5$ & $ 1706$ & $26.3$ \\
borg\_0926+4000 & borg45 & 11700 & $141.393$ & $ 40.006$ &         &        & $ 1276$ & $26.2$ & $ 2806$ & $26.3$ & $ 1106$ & $26.2$ & $  903$ & $25.9$ \\
borg\_0926+4426 &  yan28 & 11702 & $141.382$ & $ 44.426$ & $ 2374$ & $26.6$ &         &        & $ 6012$ & $27.1$ & $ 1603$ & $26.7$ & $ 1403$ & $26.5$ \\
borg\_1031+3804 & borg70 & 11700 & $157.715$ & $ 38.059$ &         &        & $ 1815$ & $26.4$ & $ 3109$ & $26.4$ & $ 1506$ & $26.3$ & $ 1306$ & $26.0$ \\
borg\_1152+5441 & borg0y & 11700 & $177.958$ & $ 54.684$ &         &        & $ 2898$ & $27.0$ & $ 6021$ & $27.0$ & $ 2809$ & $27.1$ & $ 1906$ & $26.7$ \\
borg\_1153+0056 & borg0j & 11700 & $178.182$ & $  0.931$ &         &        & $ 2647$ & $26.8$ & $ 4515$ & $26.7$ & $ 2209$ & $26.7$ & $ 1606$ & $26.4$ \\
borg\_1209+4543 & borg0p & 11700\tnm{e} & $182.355$ & $ 45.724$ & $ 2234$ & $26.6$ & $ 2707$ & $27.0$ & $13729$ & $27.4$ & $ 3709$ & $27.2$ & $ 2909$ & $26.8$ \\
borg\_1230+0750 & borg1v & 11700 & $187.470$ & $  7.825$ &         &        & $ 2436$ & $26.6$ & $ 4112$ & $26.1$ & $ 1806$ & $26.0$ & $ 1406$ & $25.6$ \\
borg\_1242+5716 &  yan11 & 11702 & $190.554$ & $ 57.270$ & $ 2800$ & $26.5$ &         &        & $ 5215$ & $27.0$ & $ 2509$ & $26.9$ & $ 2309$ & $26.6$ \\
borg\_1245+3356 & borg49 & 11700 & $191.186$ & $ 33.936$ &         &        & $ 1789$ & $26.7$ & $ 3409$ & $26.8$ & $ 1506$ & $26.7$ & $ 1106$ & $26.2$ \\
borg\_1337+0028 &  yan19 & 11702 & $204.202$ & $ -0.464$ & $ 2270$ & $26.4$ &         &        & $ 6818$ & $27.0$ & $ 1203$ & $26.6$ & $ 1203$ & $26.3$ \\
borg\_1341+4123 &  yan32 & 11702 & $205.131$ & $ 41.384$ & $ 3810$ & $27.0$ &         &        & $17435$ & $27.6$ & $ 3206$ & $27.2$ & $ 2806$ & $26.9$ \\
borg\_1437+5043 & borg58 & 11700 & $219.234$ & $ 50.719$ &         &        & $ 2754$ & $26.9$ & $ 4912$ & $27.0$ & $ 2509$ & $27.0$ & $ 1806$ & $26.5$ \\
borg\_1524+0954 &  yan51 & 11702 & $231.041$ & $  9.906$ & $ 2078$ & $26.3$ &         &        & $ 8718$ & $27.0$ & $ 1603$ & $26.5$ & $ 1303$ & $26.2$ \\
borg\_1632+3737 & borg1k & 11700 & $247.892$ & $ 37.609$ &         &        & $ 1260$ & $26.4$ & $ 2909$ & $26.7$ & $ 1206$ & $26.6$ & $  906$ & $26.0$
\enddata
\tablecomments{The total survey area of BoRG09 is $\sim135$ arcmin$^{2}$, with an effective search area for \Yband\ dropouts of $\sim98$ arcmin$^{2}$ after accounting for incompleteness.  Unless otherwise noted, each field has an area of $4.7$ arcmin$^2$.}
\tablenotetext{a}{$5\sigma$ magnitude limits in a $r=0.32\arcsec$ aperture, corrected for Galactic extinction.}
\tablenotetext{b}{Used in \cite{Trenti2011}.}
\tablenotetext{c}{Field has multiple, partially overlapping exposures for a total area $6.9$ arcmin$^2$.  Exposure times quoted are the sum of all exposures.}
\tablenotetext{d}{Data missing due to scheduling constraint/conflict.}
\tablenotetext{e}{Also includes data from HIPPIES program 11702.}
\label{tbl:borg09}
\ifemulateapj
    \end{deluxetable*}
\else
    \end{deluxetable}
\fi
}

\def\tabb {
\ifemulateapj
    \begin{deluxetable*}{llrrrrrrrrrrrr}
\else
    \begin{deluxetable}{llrrrrrrrrrrrr}
    \rotate
    \tabletypesize{\scriptsize}
\fi
\tablecolumns{14}
\tablewidth{0pt}
\tablecaption{BoRG12 Survey Fields, Exposure Times, and $5\sigma$ Limiting Magnitudes\tnm{a}}
\tablehead{ \colhead{Field} & \colhead{PID} & \colhead{$\alpha_{J2000}$}  & \colhead{$\delta_{J2000}$} & \multicolumn{2}{c}{F600LP} & \multicolumn{2}{c}{F606W} & \multicolumn{2}{c}{F098M} & \multicolumn{2}{c}{F125W} & \multicolumn{2}{c}{F160W} \\
\colhead{} & \colhead{} & \colhead{(deg)} & \colhead{(deg)} & \colhead{$t$ (s)} & \colhead{$m_{lim}$} & \colhead{$t$ (s)} & \colhead{$m_{lim}$} & \colhead{$t$ (s)} & \colhead{$m_{lim}$} & \colhead{$t$ (s)} & \colhead{$m_{lim}$} & \colhead{$t$ (s)} & \colhead{$m_{lim}$} }
\startdata
borg\_0110$-$0224\tnm{b} & 11700\tnm{d} & $ 17.532$ & $ -2.395$ & $ 3892$ & $26.7$ & $13911$ & $26.9$ & $39297$ & $27.0$ & $13538$ & $26.9$ & $10032$ & $26.6$ \\
borg\_0228$-$4102 & 11541 & $ 36.987$ & $-41.026$ & $ 1414$ & $26.5$ &         &        & $ 3406$ & $26.9$ & $ 1403$ & $26.8$ & $ 1403$ & $26.5$ \\
borg\_0240$-$1857 & 11541 & $ 40.114$ & $-18.954$ & $ 1360$ & $26.4$ &         &        & $ 3406$ & $26.8$ & $ 1403$ & $26.7$ & $ 1403$ & $26.5$ \\
borg\_0427+2538 & 11533 & $ 66.690$ & $ 25.640$ & $ 1200$ & $24.1$ &         &        & $ 2409$ & $25.2$ & $  703$ & $25.1$ & $  503$ & $25.0$ \\
borg\_0436$-$5259 & 11520 & $ 69.059$ & $-52.986$ & $ 1932$ & $26.6$ &         &        & $ 4406$ & $27.1$ & $ 2906$ & $27.3$ & $ 2206$ & $26.8$ \\
borg\_0439$-$5317 & 11520 & $ 69.855$ & $-53.278$ & $ 1932$ & $26.6$ &         &        & $ 4206$ & $27.0$ & $ 3006$ & $27.2$ & $ 2306$ & $26.8$ \\
borg\_0440$-$5244 & 11520 & $ 69.959$ & $-52.731$ & $ 1932$ & $26.5$ &         &        & $ 6609$ & $27.2$ & $ 2003$ & $27.0$ & $ 1403$ & $26.5$ \\
borg\_0835+2456 & 12025 & $128.821$ & $ 24.936$ &         &        & $ 4698$ & $26.8$ & $ 4209$ & $26.7$ & $ 2206$ & $26.7$ & $ 2009$ & $26.3$ \\
borg\_0846+7654 & 11520 & $131.593$ & $ 76.893$ & $ 1401$ & $26.4$ &         &        & $ 4406$ & $27.0$ & $ 2003$ & $27.1$ & $ 1603$ & $26.6$ \\
borg\_1010+3001 & 12025 & $152.406$ & $ 30.018$ &         &        & $ 9142$ & $27.2$ & $ 5612$ & $26.8$ & $ 3212$ & $26.8$ & $ 2812$ & $26.6$ \\
borg\_1014$-$0423 & 11524 & $153.523$ & $ -4.379$ & $ 1221$ & $26.3$ &         &        & $ 1909$ & $25.6$ & $ 1106$ & $26.2$ & $  703$ & $25.9$ \\
borg\_1031+5052\tnm{c} & 12025 & $157.691$ & $ 50.862$ &         &        & $ 9991$ & $27.4$ & $ 9629$ & $26.9$ & $ 4812$ & $27.1$ & $ 4212$ & $26.8$ \\
borg\_1033+5051\tnm{c} & 12025 & $158.212$ & $ 50.860$ &         &        & $ 9640$ & $27.3$ & $ 6412$ & $26.8$ & $ 3212$ & $26.8$ & $ 2812$ & $26.5$ \\
borg\_1051+3359 & 12024 & $162.822$ & $ 33.985$ &         &        & $ 4220$ & $26.9$ & $ 6235$ & $26.8$ & $ 3318$ & $27.0$ & $ 1912$ & $26.4$ \\
borg\_1103$-$2330 & 12025 & $165.808$ & $-23.506$ &         &        & $ 8558$ & $27.0$ & $ 9429$ & $27.1$ & $ 4412$ & $27.2$ & $ 3612$ & $26.8$ \\
borg\_1111+5545 & 12025 & $167.737$ & $ 55.751$ &         &        & $ 6524$ & $27.1$ & $ 5518$ & $26.9$ & $ 2606$ & $27.0$ & $ 2409$ & $26.6$ \\
borg\_1119+4026 & 11519 & $169.514$ & $ 40.398$ & $ 1308$ & $26.2$ &         &        & $ 2206$ & $26.6$ & $ 1403$ & $26.8$ & $ 1403$ & $26.4$ \\
borg\_1131+3114 & 11519 & $172.876$ & $ 31.289$ & $ 1316$ & $26.5$ &         &        & $ 2106$ & $26.5$ & $ 1403$ & $26.8$ & $ 1403$ & $26.4$ \\
borg\_1301+0000 & 11702 & $195.318$ & $ -0.007$ & $ 2150$ & $26.1$ &         &        & $ 5812$ & $26.4$ & $ 1603$ & $26.1$ & $ 1303$ & $25.9$ \\
borg\_1408+5503 & 12572 & $211.993$ & $ 55.056$ &         &        & $ 3324$ & $26.9$ & $ 5623$ & $26.9$ & $ 2612$ & $26.9$ & $ 2612$ & $26.6$ \\
borg\_1510+1115 & 12572 & $227.537$ & $ 11.242$ &         &        & $ 5326$ & $26.9$ & $ 8423$ & $27.1$ & $ 3812$ & $27.1$ & $ 3812$ & $26.8$ \\
borg\_1555+1108 & 12025 & $238.857$ & $ 11.132$ &         &        & $ 4753$ & $26.9$ & $ 5515$ & $27.0$ & $ 2909$ & $27.0$ & $ 2509$ & $26.7$ \\
borg\_1632+3733 & 11700 & $248.074$ & $ 37.557$ &         &        & $ 2751$ & $26.8$ & $ 5115$ & $26.9$ & $ 2406$ & $27.0$ & $ 1806$ & $26.5$ \\
borg\_1815$-$3244 & 11533 & $273.628$ & $-32.734$ & $ 1200$ & $22.3$ &         &        & $ 2509$ & $21.8$ & $  703$ & $21.6$ & $  403$ & $21.5$ \\
borg\_2057$-$4412 & 11530 & $314.340$ & $-44.207$ & $ 2500$ & $26.4$ &         &        & $ 5009$ & $26.7$ & $ 1203$ & $26.4$ & $  803$ & $25.9$ \\
borg\_2132+1004 & 11524 & $323.062$ & $ 10.064$ & $ 1355$ & $26.2$ &         &        & $ 2409$ & $26.5$ & $ 1006$ & $26.4$ & $  503$ & $25.7$ \\
borg\_2155$-$4411 & 11541 & $328.812$ & $-44.177$ & $ 2130$ & $26.7$ &         &        & $ 5609$ & $27.0$ & $ 1403$ & $26.7$ & $  903$ & $26.2$ \\
borg\_2203+1851 & 11534 & $330.705$ & $ 18.850$ & $ 3200$ & $26.8$ &         &        & $17229$ & $27.4$ & $ 2006$ & $26.8$ & $ 2806$ & $26.8$ \\
borg\_2345+0054 & 11702 & $356.261$ & $ -0.902$ & $ 2028$ & $26.4$ &         &        & $ 5612$ & $27.0$ & $ 1403$ & $26.8$ & $ 1403$ & $26.5$ \\
borg\_2351$-$4332 & 11528 & $357.650$ & $-43.525$ & $ 1050$ & $26.3$ &         &        & $11123$ & $27.4$ & $ 4209$ & $27.3$ & $ 2806$ & $26.7$
\enddata
\tablecomments{The total survey area of BoRG12 is $\sim139$ arcmin$^{2}$, with an effective search area for \Yband\ dropouts of $\sim115$ arcmin$^{2}$ after accounting for incompleteness.  The combined survey area of BoRG09+BoRG12 is $\sim 274$ arcmin$^{2}$, with an effective search area for \Yband\ dropouts of $\sim 213$ arcmin$^{2}$.  Unless otherwise noted, each field has an area of $4.7$ arcmin$^2$.}
\tablenotetext{a}{$5\sigma$ magnitude limits in a $r=0.32\arcsec$ aperture, corrected for Galactic extinction.}
\tablenotetext{b}{Field has multiple, partially overlapping exposures for a total area $14.8$ arcmin$^2$.  Exposure times quoted are the sum of all exposures.}
\tablenotetext{c}{Field has multiple, partially overlapping exposures for a total area $5.9$ arcmin$^2$.  Exposure times quoted are the sum of all exposures.}
\tablenotetext{d}{Also includes data from HIPPIES program 11702.}
\label{tbl:borg12}
\ifemulateapj
    \end{deluxetable*}
\else
    \end{deluxetable}
\fi
}

\def\tabc {
\ifemulateapj
    \begin{deluxetable*}{lrrrrrrrrr}
\else
    \begin{deluxetable}{lrrrrrrrrr}
    \rotate
    \tabletypesize{\scriptsize}
\fi
\tablecolumns{10}
\tablewidth{0pt}
\tablecaption{Photometry of \Yband-dropout ($z\sim8$) Candidates in the $8\sigma$ Catalog}
\tablehead{\colhead{ID} & \colhead{$\alpha_{J2000}$} & \colhead{$\delta_{J2000}$} & \colhead{\Jband\tnm{a}} & \colhead{\Yband $-$ \Jband} & \colhead{\Jband $-$ \Hband} & \colhead{$S/N_{V}$\tnm{b}} & \colhead{$S/N_{098}$} & \colhead{$S/N_{125}$} & \colhead{$S/N_{160}$} }
\startdata
borg\_0440$-$5244\_682  & $69.9455843$  & $-52.7320162$ & $25.9 \pm  0.1$ & $> 2.1$         & $ 0.0 \pm  0.3$ & $-0.8$ & $-0.5$ & $ 9.1$ & $ 5.7$ \\
borg\_0751$+$2917\_229  & $117.7141714$ & $29.2715323$  & $26.5 \pm  0.2$ & $> 2.5$         & $-0.0 \pm  0.3$ & $-0.9$ & $ 0.0$ & $ 8.9$ & $ 6.5$ \\
borg\_0909$+$0002\_595  & $137.2731625$ & $-0.0297391$  & $26.1 \pm  0.2$ & $ 1.8 \pm  0.6$ & $-0.4 \pm  0.4$ & $-0.5$ & $ 1.8$ & $ 8.6$ & $ 3.2$ \\
borg\_1033$+$5051\_126  & $158.1863098$ & $50.8416866$  & $26.0 \pm  0.2$ & $> 2.5$         & $-0.2 \pm  0.3$ & $ 0.0$ & $-0.6$ & $ 8.1$ & $ 5.6$ \\
borg\_1301$+$0000\_160  & $195.3070838$ & $-0.0189297$  & $25.5 \pm  0.2$ & $ 2.1 \pm  0.6$ & $-0.4 \pm  0.3$ & $ 0.7$ & $ 1.7$ & $ 9.5$ & $ 5.2$ \\
borg\_1437$+$5043\_1137 & $219.210672$  & $50.7260085$  & $26.1 \pm  0.1$ & $> 2.7$         & $ 0.0 \pm  0.2$ & $-1.5$ & $-1.0$ & $10.9$ & $ 7.9$ \\
borg\_1555$+$1108\_1417 & $238.8651549$ & $11.1393576$  & $26.6 \pm  0.2$ & $ 1.9 \pm  0.7$ & $-0.6 \pm  0.4$ & $ 1.1$ & $ 1.4$ & $ 8.4$ & $ 3.4$ \\
borg\_2203$+$1851\_1061 & $330.6930404$ & $18.8581986$  & $26.0 \pm  0.2$ & $ 2.5 \pm  0.6$ & $-0.3 \pm  0.3$ & $ 0.2$ & $ 1.6$ & $ 9.4$ & $ 6.6$
\enddata
\tablenotetext{a}{Total magnitudes (AUTOMAG).}
\tablenotetext{b}{$V_{606}$ whenever possible; otherwise $V_{600LP}$.}
\label{tbl:8sig}
\ifemulateapj
    \end{deluxetable*}
\else
    \end{deluxetable}
\fi
}

\def\tabd {
\ifemulateapj
    \begin{deluxetable*}{lrrrrrrrrr}
\else
    \begin{deluxetable}{lrrrrrrrrr}
    \rotate
    \tabletypesize{\scriptsize}
\fi
\tablecolumns{10}
\tablewidth{0pt}
\tablecaption{Photometry of \Yband-dropout ($z\sim8$) Candidates in the $5\sigma$ Catalog}
\tablehead{\colhead{ID} & \colhead{$\alpha_{J2000}$} & \colhead{$\delta_{J2000}$} & \colhead{\Jband\tnm{a}} & \colhead{\Yband $-$ \Jband} & \colhead{\Jband $-$ \Hband} & \colhead{$S/N_{V}$\tnm{b}} & \colhead{$S/N_{098}$} & \colhead{$S/N_{125}$} & \colhead{$S/N_{160}$} }
\startdata
      borg\_0436$-$5259\_1233 & $69.0303878$ & $-52.9717897$ & $27.1 \pm  0.2$ & $> 1.8$ & $-0.4 \pm  0.5$ & $ 1.4$ & $ 0.9$ & $ 5.5$ & $ 2.6$ \\
      borg\_0553$-$6405\_4006 & $88.2647181$ & $-64.0821631$ & $26.7 \pm  0.2$ & $> 2.3$ & $ 0.1 \pm  0.3$ & $ 0.2$ & $ 0.6$ & $ 6.8$ & $ 4.7$ \\
      borg\_0751$+$2917\_920 & $117.706444$ & $29.2977181$ & $27.0 \pm  0.2$ & $> 2.3$ & $-0.6 \pm  0.4$ & $-0.3$ & $-0.0$ & $ 7.1$ & $ 3.3$ \\
      borg\_0756$+$3043\_437 & $118.9794155$ & $30.7177854$ & $26.5 \pm  0.2$ & $> 2.2$ & $-0.0 \pm  0.3$ & $ 1.3$ & $ 0.9$ & $ 6.8$ & $ 4.6$ \\
      borg\_0835$+$2456\_253 & $128.8067291$ & $24.9267442$ & $26.2 \pm  0.2$ & $ 2.0 \pm  0.8$ & $-0.2 \pm  0.4$ & $-1.2$ & $ 1.1$ & $ 6.8$ & $ 4.1$ \\
      borg\_1031$+$3804\_213 & $157.7102797$ & $38.0497128$ & $26.6 \pm  0.3$ & $> 1.9$ & $-0.1 \pm  0.4$ & $ 0.3$ & $ 0.7$ & $ 5.4$ & $ 4.0$ \\
      borg\_1031$+$3804\_831 & $157.7353387$ & $38.0673682$ & $26.6 \pm  0.3$ & $> 1.9$ & $-0.5 \pm  0.5$ & $ 1.4$ & $ 0.3$ & $ 5.5$ & $ 2.9$ \\
      borg\_1103$-$2330\_1180 & $165.7881655$ & $-23.4990798$ & $26.7 \pm  0.2$ & $> 2.2$ & $ 0.0 \pm  0.3$ & $-0.3$ & $-0.4$ & $ 7.8$ & $ 5.8$ \\
      borg\_1131$+$3114\_1244 & $172.8573532$ & $31.2942314$ & $26.2 \pm  0.2$ & $> 2.0$ & $-0.3 \pm  0.3$ & $ 0.8$ & $-0.2$ & $ 7.7$ & $ 4.3$ \\
      borg\_1152$+$5441\_1087 & $177.9750532$ & $54.6979452$ & $27.2 \pm  0.3$ & $> 1.8$ & $-0.2 \pm  0.4$ & $ 1.2$ & $ 0.9$ & $ 5.5$ & $ 3.2$ \\
      borg\_1153$+$0056\_540 & $178.1909931$ & $0.9320074$ & $27.0 \pm  0.3$ & $ 1.9 \pm  0.8$ & $-0.5 \pm  0.4$ & $ 0.2$ & $ 1.1$ & $ 5.9$ & $ 3.2$ \\
      borg\_1242$+$5716\_159 & $190.5672023$ & $57.2567197$ & $26.4 \pm  0.2$ & $> 2.2$ & $-0.2 \pm  0.4$ & $ 0.0$ & $ 0.4$ & $ 6.6$ & $ 4.3$ \\
      borg\_1408$+$5503\_749 & $212.0126402$ & $55.0585147$ & $26.5 \pm  0.2$ & $> 2.0$ & $-0.2 \pm  0.3$ & $-1.1$ & $ 0.8$ & $ 6.7$ & $ 4.4$ \\
      borg\_1408$+$5503\_980 & $212.0082405$ & $55.0672755$ & $27.0 \pm  0.2$ & $ 1.9 \pm  0.8$ & $-0.4 \pm  0.4$ & $-0.2$ & $ 1.0$ & $ 6.0$ & $ 3.5$ \\
      borg\_1437$+$5043\_172 & $219.2223469$ & $50.7080907$ & $27.1 \pm  0.2$ & $ 2.0 \pm  0.8$ & $-0.4 \pm  0.5$ & $ 1.0$ & $ 1.0$ & $ 5.8$ & $ 2.8$ \\
      borg\_1437$+$5043\_879 & $219.2240496$ & $50.7259683$ & $27.3 \pm  0.3$ & $> 1.8$ & $-0.3 \pm  0.5$ & $ 0.3$ & $ 0.5$ & $ 5.0$ & $ 2.7$ \\
      borg\_1510$+$1115\_51 & $227.5348783$ & $11.2225448$ & $26.7 \pm  0.2$ & $ 1.9 \pm  0.8$ & $-0.2 \pm  0.4$ & $ 1.2$ & $ 1.0$ & $ 6.2$ & $ 3.9$ \\
      borg\_1510$+$1115\_1236 & $227.5521577$ & $11.2522441$ & $27.2 \pm  0.3$ & $> 2.0$ & $-0.4 \pm  0.4$ & $ 1.1$ & $ 0.7$ & $ 6.3$ & $ 3.5$ \\
      borg\_1510$+$1115\_1404 & $227.5425951$ & $11.2615405$ & $26.7 \pm  0.2$ & $> 1.9$ & $-0.3 \pm  0.4$ & $ 0.8$ & $ 0.7$ & $ 6.5$ & $ 4.0$ \\
      borg\_1555$+$1108\_595 & $238.8429322$ & $11.1279017$ & $27.3 \pm  0.3$ & $> 1.9$ & $-0.2 \pm  0.4$ & $ 0.8$ & $-0.1$ & $ 5.7$ & $ 3.2$ \\
      borg\_1555$+$1108\_1166 & $238.8438689$ & $11.1420588$ & $27.2 \pm  0.2$ & $> 1.9$ & $-0.3 \pm  0.4$ & $ 1.2$ & $ 0.2$ & $ 5.8$ & $ 2.9$ \\
      borg\_1632$+$3733\_694 & $248.0628393$ & $37.5568592$ & $27.4 \pm  0.3$ & $> 2.0$ & $-0.4 \pm  0.5$ & $ 0.9$ & $ 0.7$ & $ 6.2$ & $ 2.7$ \\
      borg\_2132$+$1004\_24 & $323.0575947$ & $10.0443562$ & $26.5 \pm  0.2$ & $> 2.1$ & $ 0.0 \pm  0.4$ & $ 1.4$ & $ 0.6$ & $ 5.3$ & $ 3.2$ \\
      borg\_2155$-$4411\_341 & $328.8301017$ & $-44.1819169$ & $26.6 \pm  0.2$ & $> 2.5$ & $-0.4 \pm  0.4$ & $ 0.0$ & $-1.1$ & $ 7.3$ & $ 3.4$ \\
      borg\_2155$-$4411\_1192 & $328.8031162$ & $-44.1737506$ & $26.9 \pm 0.3$ & $> 2.1$ & $-0.1 \pm 0.5$ & $-0.7$ & $ 0.6$ & $ 5.2$ & $ 3.1$
\enddata
\tablecomments{Sources reported in the $8\sigma$ catalog (Table~\ref{tbl:8sig}) are not duplicated here, but of course are included in the $5\sigma$ catalog.}
\tablenotetext{a}{Total magnitudes (AUTOMAG).}
\tablenotetext{b}{$V_{606}$ whenever possible; otherwise $V_{600LP}$.}
\label{tbl:5sig}
\ifemulateapj
    \end{deluxetable*}
\else
    \end{deluxetable}
\fi
}

\def\tabe {
\ifemulateapj
    \begin{deluxetable*}{lrrrrrrrrr}
\else
    \begin{deluxetable}{lrrrrrrrrr}
    \rotate
    \tabletypesize{\scriptsize}
\fi
\tablecolumns{10}
\tablewidth{0pt}
\tablecaption{New Photometry of \Yband-dropout ($z\sim8$) Candidates in the BoRG58 Protocluster \citep{Trenti2012a}}
\tablehead{\colhead{ID} & \colhead{$\alpha_{J2000}$} & \colhead{$\delta_{J2000}$} & \colhead{\Jband\tnm{a}} & \colhead{\Yband $-$ \Jband} & \colhead{\Jband $-$ \Hband} & \colhead{$S/N_{606}$} & \colhead{$S/N_{098}$} & \colhead{$S/N_{125}$} & \colhead{$S/N_{160}$} }
\startdata
borg\_1437$+$5043\_1137 & $219.210672$ & $50.7260085$ & $26.1 \pm  0.1$ & $> 2.7$        & $ 0.0 \pm  0.2$ & $-1.5$ & $-1.0$ & $10.9$ & $ 7.9$ \\
borg\_1437$+$5043\_879 & $219.2240496$ & $50.7259683$ & $27.3 \pm  0.3$ & $> 1.8$        & $-0.3 \pm  0.5$ & $ 0.3$ & $ 0.5$ & $ 5.0$ & $ 2.7$ \\
borg\_1437$+$5043\_757 & $219.2310489$ & $50.7240585$ & $27.1 \pm  0.2$ & $> 1.8$        & $-0.7 \pm  0.6$ & $ 1.1$ & $-0.5$ & $ 5.0$ & $ 1.7$ \\
borg\_1437$+$5043\_435 & $219.2202746$ & $50.7156344$ & $27.4 \pm  0.3$ & $> 1.8$        & $ 0.0 \pm  0.4$ & $ 0.6$ & $ 0.6$ & $ 4.9$ & $ 3.3$ \\
borg\_1437$+$5043\_172 & $219.2223469$ & $50.7080907$ & $27.1 \pm  0.2$ & $ 2.0 \pm 0.8$ & $-0.4 \pm  0.5$ & $ 1.0$ & $ 1.0$ & $ 5.8$ & $ 2.8$
\enddata
\tablenotetext{a}{Total magnitudes (AUTOMAG).}
\label{tbl:borg58}
\ifemulateapj
    \end{deluxetable*}
\else
    \end{deluxetable}
\fi
}

\def\tabf {
\ifemulateapj
    \begin{deluxetable*}{lrrrrrrrrr}
\else
    \begin{deluxetable}{lrrrrrrrrr}
    \rotate
    \tabletypesize{\scriptsize}
\fi
\tablecolumns{10}
\tablewidth{0pt}
\tablecaption{Photometry of Other Possible \Yband-dropout ($z\sim8$) Candidates}
\tablehead{\colhead{ID} & \colhead{$\alpha_{J2000}$} & \colhead{$\delta_{J2000}$} & \colhead{\Jband\tnm{a}} & \colhead{\Yband $-$ \Jband} & \colhead{\Jband $-$ \Hband} & \colhead{$S/N_{V}$\tnm{b}} & \colhead{$S/N_{098}$} & \colhead{$S/N_{125}$} & \colhead{$S/N_{160}$} }
\startdata
borg\_0240$-$1857\_392\tnm{c} & $40.0998766$ & $-18.9604896$ & $26.2 \pm  0.2$ & $ 2.1 \pm  0.7$ & $-0.1 \pm  0.3$ & $ 0.6$ & $ 1.4$ & $ 8.1$ & $ 6.3$ \\
borg\_1632$+$3737\_386\tnm{d} & $247.8986483$ & $37.6047539$ & $25.1 \pm  0.1$ & $ 1.9 \pm  0.4$ & $ 0.0 \pm  0.2$ & $-0.0$ & $ 3.1$ & $14.2$ & $ 9.3$
\enddata
\tablenotetext{a}{Total magnitudes (AUTOMAG).}
\tablenotetext{b}{$V_{606}$ whenever possible; otherwise $V_{600LP}$.}
\tablenotetext{c}{Unresolved source; possible QSO candidate or most likely an L/T dwarf star with an unusually red $Y_{098} - J_{125}$ color.}
\tablenotetext{d}{Photometry contaminated by an adjacent bright source with similar colors ($Y_{098} - J_{125} = 1.6$), but clearly detected in \Vband\ and \Yband; most likely a low-redshift contaminant.}
\label{tbl:other}
\ifemulateapj
    \end{deluxetable*}
\else
    \end{deluxetable}
\fi
}

\def\tabg {
\ifemulateapj
    \begin{deluxetable}{cc}
\else
    \begin{deluxetable}{cc}
\fi
\tablecolumns{2}
\tablewidth{0pt}
\tablecaption{BoRG Stepwise Determination of the $z \sim 8$ UV LF\tnm{a}}
\tablehead{\colhead{$M_{UV}$} & \colhead{$\phi_{k}$ ($10^{-4}$~Mpc$^{-3}$ mag$^{-1}$)\tnm{b}} }
\startdata
$-22.14$  &  $<0.015$                    \\
$-21.64$  &  $0.023^{+0.053}_{-0.019}$   \\
$-21.14$  &  $0.198^{+0.106}_{-0.072}$   \\
$-20.64$  &  $0.604^{+0.217}_{-0.163}$   \\
$-20.14$  &  $1.296^{+0.591}_{-0.418}$   \\
$-19.64$  &  $4.504^{+4.366}_{-2.434}$
\enddata
\tablecomments{Assuming $h = 0.7$.}
\tablenotetext{a}{For the $5\sigma$ sample.}
\tablenotetext{b}{The errors are derived from the 68\% Bayesian credible intervals for a Poisson distribution.}
\label{tbl:swlf}
\ifemulateapj
    \end{deluxetable}
\else
    \end{deluxetable}
\fi
}

\def\tabh {
\ifemulateapj
    \begin{deluxetable*}{crcrr}
\else
    \begin{deluxetable}{crcrr}
\fi
\tablecolumns{5}
\tablewidth{0pt}
\tablecaption{Effect of the Contamination Fraction on the LF Determination}
\tablehead{\colhead{$1/(1-f)$} & \colhead{$\ln \mathcal{L}$} & \colhead{$\phi_{*} (10^{-4}$~Mpc$^{-3})$} & \colhead{$M_{*}$} & \colhead{$\alpha$} }
\startdata
$1.0$         & $-24.56$ & $5.6$ & $-20.28$ & $-1.80$ \\
$1.25$        & $-23.96$ & $4.7$ & $-20.31$ & $-1.90$ \\
$1.5$         & $-23.84$ & $4.4$ & $-20.29$ & $-1.95$ \\
$1.73$\tnm{a} & $-23.98$ & $4.3$ & $-20.26$ & $-1.97$ \\
$2.0$         & $-24.35$ & $3.8$ & $-20.28$ & $-2.05$ \\
$2.25$        & $-24.80$ & $3.5$ & $-20.28$ & $-2.10$ \\
$2.5$         & $-25.31$ & $3.6$ & $-20.23$ & $-2.10$
\enddata
\tablecomments{Best-fitting parameters (columns 3-5) of the $z\sim8$ LF as a function of the contamination correction applied (column 1) to the BoRG $5\sigma$ sample, with the likelihood $\ln \mathcal{L}$ for the fit shown in the second column.}
\tablecomments{Assuming $h = 0.7$.}
\tablenotetext{a}{Our fiducial value for the contamination correction.}
\label{tbl:contam}
\ifemulateapj
    \end{deluxetable*}
\else
    \end{deluxetable}
\fi
}

\def\tabi {
\ifemulateapj
    \begin{deluxetable*}{lccc}
\else
    \begin{deluxetable}{lccc}
\fi
\tablecolumns{4}
\tablewidth{0pt}
\tablecaption{Comparison of $z\sim8$ LF Determinations in the Literature}
\tablehead{\colhead{Reference} & \colhead{$\log \phi_{*}$ (Mpc$^{-3}$)} & \colhead{$M_{*}$} & $\alpha$ }
\startdata
This Work               & $-3.37^{+0.26}_{-0.29}$  & $-20.26^{+0.29}_{-0.34}$   & $-1.98^{+0.23}_{-0.22}$  \\
\citet{Oesch2012}       & $-3.17^{+0.40}_{-0.55}$  & $-19.80 ^{+0.46}_{-0.57}$  & $-2.06^{+0.45}_{-0.37}$  \\
\citet{Bouwens2011b}    & $-3.23^{+0.74}_{-0.27}$  & $-20.10\pm0.52$            & $-1.91\pm0.32$           \\
\citet{Lorenzoni2011}   & $-3.0$                   & $-19.5$                    & $-1.7$ (fixed)           \\
\citet{Trenti2011}      & $-3.4$ (fixed)           & $ -20.2\pm0.3$             & $-2.0$ (fixed)           \\
\citet{McLure2010}      & $-3.46$                  & $-20.04$ (fixed)           & $-1.71$ (fixed)          \\
\citet{Bouwens2010a}    & $-2.96$ (fixed)          & $-19.5\pm0.3$              & $-1.74$ (fixed)
\enddata
\label{tbl:lfparam}
\tablecomments{Assuming $h = 0.7$.}
\ifemulateapj
    \end{deluxetable*}
\else
    \end{deluxetable}
\fi
}

\ifemulateapj
    \journalinfo{Accepted to The Astrophysical Journal}
    \slugcomment{Accepted to The Astrophysical Journal}
    \shorttitle{BoRG: The Bright End of the $z\sim8$ LF}
    \shortauthors{BRADLEY ET AL.}
\fi

\begin{document}

\title{The Brightest of Reionizing Galaxies Survey:  Constraints on the Bright End of the \lowercase{$z\sim8$} Luminosity Function\altaffilmark{*}}

\author{L.D.~Bradley\altaffilmark{1}, M.~Trenti\altaffilmark{2,3,\dag}, P.A.~Oesch\altaffilmark{4,\ddag}, M.~Stiavelli\altaffilmark{1}, T.~Treu\altaffilmark{5}, R.J.~Bouwens\altaffilmark{6}, J.M.~Shull\altaffilmark{7}, B.W.~Holwerda\altaffilmark{8}, N.~Pirzkal\altaffilmark{1}}

\altaffiltext{*}{Based on observations made with the NASA/ESA {\em Hubble Space Telescope}, obtained at the Space Telescope Science Institute, which is operated by the Association of Universities for Research in Astronomy under NASA contract NAS5-26555.  These observations are associated with programs 11519, 11520, 11524, 11528, 11530, 11533, 11534, 11541, 11700, 11702, 12024, 12025, and 12572.}
\altaffiltext{1}{Space Telescope Science Institute, 3700 San Martin Drive Baltimore MD 21218 USA}
\altaffiltext{2}{Institute of Astronomy, University of  Cambridge, Madingley Road, Cambridge,  CB3 0HA, United Kingdom}
\altaffiltext{3}{Kavli Institute for Cosmology, University of  Cambridge, Madingley Road, Cambridge,  CB3 0HA, United Kingdom}
\altaffiltext{4}{UCO/Lick Observatory, University of California, Santa Cruz, CA 95064 USA}
\altaffiltext{5}{Department of Physics, University of California, Santa Barbara, CA 93106-9530, USA}
\altaffiltext{6}{Leiden Observatory, Leiden University, NL-2300 RA Leiden, The Netherlands}
\altaffiltext{7}{CASA, Department of Astrophysical and Planetary Sciences, University of Colorado, 389-UCB, Boulder, CO 80309 USA}
\altaffiltext{8}{European Space Agency (ESTEC), Keplerlaan 1, 2200 AG, Noordwijk, The Netherlands}
\altaffiltext{\dag}{Kavli Fellow}
\altaffiltext{\ddag}{Hubble Fellow}

\begin{abstract}
We report the discovery of 33 Lyman break galaxy (LBG) candidates at
$z\sim 8$ detected in {\em Hubble Space Telescope (HST)} Wide Field
Camera 3 (WFC3) imaging as part of the Brightest of Reionizing
Galaxies (BoRG) pure-parallel survey.  The ongoing BoRG survey
currently has the largest area ($274$ arcmin$^2$) with \Yband\ (or
$Y_{105}$), \Jband, and \Hband\ band coverage needed to search for
$z\sim~8$ galaxies, about three times the current CANDELS area, and
slightly larger than what will be the final CANDELS wide component
with $Y_{105}$ data (required to select $z\sim8$ sources).  Our sample
of 33 relatively bright \Yband-dropout galaxies have \Jband\ band
magnitudes between $25.5$ and $27.4$ mag.  This is the largest sample
of bright ($\Jband \la 27.4$) $z\sim8$ galaxy candidates presented to
date.  Combining our dataset with the Hubble Ultra-Deep Field (HUDF09)
dataset, we constrain the rest-frame ultraviolet galaxy luminosity
function at $z\sim8$ over the widest dynamic range currently
available.  The combined datasets are well fitted by a Schechter
function, i.e. $\phi(L) = \phi_{*} (L/L_{*})^{\alpha}\
e^{-(L/L_{*})}$, without evidence for an excess of sources at the
bright end.  At $68\%$ confidence, for $h=0.7$ we derive $\phi_{*} =
(4.3^{+3.5}_{-2.1}) \times 10^{-4}$ Mpc$^{-3}$, $\Mstar =
-20.26^{+0.29}_{-0.34}$, and a very steep faint-end slope $\alpha =
-1.98^{+0.23}_{-0.22}$.  While the best-fit parameters still have a
strong degeneracy, especially between $\phi_{*}$ and $M_{*}$, our
improved coverage at the bright end has reduced the uncertainty of the
faint-end power-law slope at $z\sim8$ compared to the best previous
determination at $\pm 0.4$.  With a future expansion of the BoRG
survey, combined with planned ultradeep WFC3/IR observations, it will
be possible to further reduce this uncertainty and clearly demonstrate
the steepening of the faint-end slope compared to measurements at
lower redshift, thereby confirming the key role played by small
galaxies in the reionization of the universe.
\end{abstract}


\section{Introduction}

Finding the earliest galaxies in the universe and characterizing their
properties and contribution to the reionization of the universe are
some of the most import goals of extragalactic astronomy.  The Wide
Field Camera 3 (WFC3) aboard the {\em Hubble Space Telescope} (\HST)
has significantly expanded the high-redshift frontier with the
detection and study of galaxies at $z\ga7$.  The ultradeep WFC3/IR
observations of the Hubble Ultra-Deep Field (HUDF09) and nearby fields
have so far yielded 73 $z\sim7$ and 59 $z\sim8$ Lyman-break galaxy
(LBG) candidates \citep[][see also \citealt{Lorenzoni2011,
McLure2011}]{Bouwens2011b}, including one at $z\sim10$
\citep{Bouwens2011a}.  These ultradeep observations show tantalizing
evidence for a rapid evolution of the rest-frame ultraviolet (UV)
galaxy luminosity function (LF) from $z = 6$ to $z = 8$ and a
declining star-formation rate with increasing redshift
\citep{Bouwens2011a, Bouwens2011b}, as expected on the basis of dark
matter halo assembly \citep{Trenti2010}.

However, because the ultradeep datasets cover only a few WFC3/IR
fields, their dynamic range is limited by the small volume that they
probe.  In particular they provide poor constraints on the population
of bright galaxies at $z\sim8$, which are very rare and highly
clustered \citep{Trenti2012a}.  A complete understanding of the number
density and overall shape of the LF at $z\sim8$ requires a large-area
survey to search for the brightest galaxies at these epochs.

Identifying the brightest $z\sim8$ candidates from broad-band
photometry is also of fundamental importance to provide the best
targets for spectroscopic follow-up studies aimed both at confirming
the redshift of the sources as well as inferring the properties of the
intergalactic medium (IGM) in proximity of these galaxies
\citep[e.g.,][]{Schenker2012, Treu2012}.  Spectroscopic confirmation
of the redshift of LBG galaxies has been carried out for large samples
at $z\sim 4-6$ \citep[e.g.,][]{Malhotra2005, Stark2011}, and recently
extended out to $z=7.2$ \citep{Ono2012}.  These studies show that
photometrically selected samples of LBGs have very low contamination
($\sim 10\%$, see \citealt{Malhotra2005}).  It is necessary to extend
the frontier of spectroscopy further into the epoch of reionization,
at $z\sim 8$, not only to provide definitive proof that the LBG
selection continues to be reliable for candidates into the
reionization epoch, but more importantly to infer the ionization state
of the IGM from the study of the Ly$\alpha$ equivalent width
distribution of LBG sources \citep[e.g.,][]{Treu2012}.

At present there are several ongoing WFC3/IR surveys designed to
identify a relatively large sample of bright ($m_{AB} \la 27$; $M_{AB}
\la -20$) $z\sim8$ galaxies, namely the CANDELS Multi-Cycle Treasury
(MCT) program \citep{Grogin2011, Koekemoer2011}, the Hubble Infrared
Pure Parallel Imaging Extragalactic Survey (HIPPIES) \citep{Yan2011a},
and the Brightest of Reionizing Galaxies (BoRG) \citep{Trenti2011,
Trenti2012a}.  In the near term, the sample size of bright $z\ga7$
galaxies will also be augmented by cluster lensing surveys
\citep[e.g.,][]{Hall2012, Bradley2012}, including the Cluster Lensing
And Supernova survey with Hubble (CLASH; \citealt{Postman2012}) MCT
program, which recently reported the discovery of a magnified LBG
candidate at $z\sim9.6$ \citep{Zheng2012}.  All of these programs take
advantage of multi-band \HST\ optical NIR data to search for $z\sim8$
galaxies as \Yband\ or $Y_{105}$ dropouts using the Lyman-break
technique \citep{Steidel1996a}.

Currently, the shape of the bright-end of the galaxy UV LF at $z=8$ is
debated.  Theoretical and numerical investigations predict that it
should remain Schechter-like \citep{Schechter1976}, of the form
$\phi(L) = \phi_{*} (L/L_{*})^{\alpha}\ e^{-(L/L_{*})}$, at
$z\sim7-10$ \citep{Trenti2010, Jaacks2012}.  Conversely, some
numerical studies have suggested a possible excess of sources at the
bright end, with a non-Schechter behavior tied to the inefficient
onset of AGN feedback at early times \citep[e.g.,][]{Finlator2011}.

Observationally, two recent papers from the CANDELS dataset in
GOODS-South yielded different results.  \citet{Yan2012} found a
significant excess of bright $z\sim8$ candidates with $M_{AB} < -21.0$
over the expectation from a Schechter function using data from
$\sim80$~arcmin${^2}$ of area in the GOODS-South field.  The shape of
their $z\sim8$ luminosity function resembles a step function (see
their Figure 4).  On the other hand, \citet{Oesch2012} analyzed the
same dataset and derive a well-behaved Schechter function for bright
$z\sim8$ sources, consistent with the extrapolation of the LF
evolution from $z\sim7$ and with the predictions of the LF model of
\citet{Trenti2010}.  The \citet{Oesch2012} result ($M^{*}_{z=8} =
-19.8^{+0.46}_{-0.57}$) is also consistent with the first-epoch BoRG
(BoRG09) determination of the knee of the Schechter function
($M^{*}_{z=8} = -20.2\pm0.3$, with the error bar at a fixed $\alpha$
and $\phi_{*}$).  Our initial BoRG data were about $0.5$ mag shallower
than the current CANDELS dataset, but we already had larger sky
coverage ($\sim140$~arcmin$^{2}$ vs. $\sim80$~arcmin$^{2}$).

Here we take advantage of the additional fields recently acquired as
part of the BoRG survey and update the sample of \Yband\ dropouts
described in \citet{Trenti2011, Trenti2012a} to include an additional
139 arcmin$^2$ of area.  Our latest catalog now contains eight very
bright ($>8\sigma$) $z\sim8$ candidates and an additional 25 bright
$z\sim8$ candidates detected at lower significance ($>5\sigma$), with
$M\sim M^{*}_{z=8}$.  This is currently the largest-area search for
$z\sim8$ candidates, now totaling $\sim274$ arcmin$^{2}$.
Additionally, given the random pointing nature of our pure-parallel
\HST\ program, the BoRG dataset is distinct in that it is minimally
affected by cosmic variance \citep{Trenti2008}.  Therefore, this
catalog is uniquely positioned to set the tightest constraints on the
number density of the brightest $z\sim8$ galaxies and the bright end
of the rest-frame ultraviolet galaxy luminosity function ($M_{UV} \le
-19.6$; $L \ge 0.5 L^{*}_{z=8}$).

This paper is organized as follows.  We describe the BoRG survey in
Section~\ref{sect:survey}.  We present the observations and photometry
in Section~\ref{sect:obs} and dropout selection in
Section~\ref{sect:select}.  The results and constraints on the
$z\sim8$ UV LF are discussed in Section~\ref{sect:uvlf}.  We summarize
our results and conclusions in Section~\ref{sect:concl}.  Throughout
this work, we assume a cosmology with $\Omega_{m} = 0.3$,
$\Omega_{\Lambda} = 0.7$, and $H_{0} = 70$~\kmsMpc.  This provides an
angular scale of 4.8 (proper)~kpc~\peras\ at $z = 8.0$.  All
magnitudes are expressed in the AB photometric system \citep{Oke1974}.

\ifemulateapj\figa\fi
\ifemulateapj\taba\fi
\ifemulateapj\tabb\fi
\ifemulateapj\tabc\fi
\ifemulateapj\tabd\fi
\ifemulateapj\tabe\fi
\ifemulateapj\tabf\fi

\section{The BoRG Survey for \lowercase{$z\sim8$}
Galaxies}\label{sect:survey}

In 2009 we initiated the BoRG survey, a pure-parallel WFC3 survey that
complements deep and ultradeep WFC3/IR observations by looking for
very bright ($m_{AB} \la 27$; $M_{AB} \la -20$) galaxies at $z\ga7.5$
\citep{Trenti2011, Trenti2012a} by obtaining WFC3 imaging in four
filters (F606W, F098M, F125W, F160W) on random sightlines at high
Galactic latitudes ($|b| > 30\dg$).  Because luminous massive galaxies
at these redshifts are expected to be clustered \citep{Trenti2012a},
the random pointing nature of our BoRG survey is ideal to mitigate the
severe effects of cosmic variance \citep{Trenti2008, Robertson2010a}.
In fact, our current survey geometry of 59 independent WFC3/IR fields
makes our number counts of $z\sim8$ galaxies essentially follow a
Poisson distribution and can be used to constrain the luminosity
function as well as a contiguous single-field survey with about two
times more area at the same depth.  If all the BoRG area had been in a
single field of $16.5 \times 16.5$ arcmin$^2$, we derive from
\citet{Trenti2008} that cosmic variance would have dominated over
Poisson uncertainty for a sample of 33 candidates ($22\%$ vs. $17\%$
respectively, for a total fractional error of $28\%$).

By reducing the uncertainty in the number density of bright sources
due to cosmic variance, we can place stronger observational
constraints on \Lstar\ (or \Mstar), the characteristic luminosity of
the LF.  This, in turn, can help break the well-known degeneracy
between \Mstar\ and $\alpha$, the faint-end slope, when fitting a
Schechter \citeyearpar{Schechter1976} luminosity function to the data.
Placing tighter constraints on the value of $\alpha$ at redshifts
$z\ga6$ is crucial in determining the contribution of galaxies to the
reionization of the universe.  Further, the relative brightness of the
LBGs discovered in the BoRG survey can possibly place $z\sim8$
galaxies within reach of ground-based spectroscopic follow-up
observations, as has been attempted on some of our brighter candidates
\citep{Schenker2012, Treu2012}.

Our ongoing BoRG survey has covered 274 arcmin$^{2}$ to date thanks
both to its continuation in Cycle 19 (GO 12752, PI Trenti) and by
assimilating data from the similar HIPPIES pure-parallel program
\citep{Yan2011a} as well as the coordinated parallel observations
acquired as part of the Cosmic Origins Spectrograph (COS) GTO program.
As illustrated in Figure~\ref{fig:survey}, the BoRG survey is
$\sim0.3$ mag shallower than the ``wide'' part of the CANDELS program
that has $Y_{105}$-band data, but has a larger area with \Yband\ (or
$Y_{105}$), \Jband, and \Hband\ band coverage that is needed to search
for $z\sim8$ galaxies.

In this paper we focus on $z\sim8$ LBG candidates selected as
\Yband-band dropouts.  To use a homogeneous selection of $z\sim8$
candidates, we specifically exclude the HIPPIES Cycle 18 dataset,
which replaced the F098M band with the F105W band.  We leave the
identification of $z\sim8$ $Y_{105}$-band dropouts for discussion in a
future paper.

\ifemulateapj\figb\fi
\ifemulateapj\figc\fi
\ifemulateapj\figd\fi

\section{Observations and Photometry}\label{sect:obs}

The primary goal of the BoRG survey is to identify galaxies at
$z\sim8$ and measure the bright end of the LBG rest-frame ultraviolet
LF at this redshift.  Our observations are designed to acquire WFC3
imaging in four filters (F606W, F098M, F125W, F160W), which are
obtained on random and discrete sightlines inherent to this being a
pure-parallel program.  We identify bright ($\Jband \la 27$)
high-redshift galaxies by searching for \Yband\ dropouts using the
well-known Lyman-break dropout technique \citep{Steidel1996a}.  The
\Jband\ and \Hband\ bands are used for source detection and to measure
the rest-frame UV color, while the WFC3/UVIS \Vband\ band is used to
reject low-redshift ($z\sim1.5-2$) interloper galaxies, which are the
primary source of contamination to our $z\ga7.5$ sample.  As discussed
in Section~\ref{sect:select}, our BoRG survey is optimized to minimize
the probability of contamination from both low-redshift interlopers
and cool dwarf stars.

We also included data from the similar HIPPIES WFC3 pure-parallel
program \cite{Yan2011a} and the coordinated parallel observations from
the COS GTO program.  The filter selection of the HIPPIES program and
the earlier part of COS GTO program differs from BoRG in that those
datasets used the F600LP filter instead of F606W to control for
contamination from reddened low-redshift sources.  \citet{Trenti2011}
discuss the benefits of F606W compared to F600LP.  Here we reiterate
that, because F606W has a larger transmission efficiency integrated in
frequency space over the passband, it reaches deeper at fixed
integration time.  This also becomes apparent in the data, for example
in field borg\_$0751+2917$ (Table~\ref{tbl:borg09}), which has deeper
data in F606W ($m_{lim}=26.8$ with $t=2826$ s) compared to F600LP
($m_{lim}=26.6$ with $t=3732$ s) despite the shorter integration time.

Pure-parallel observations come with a unique set of challenges.
Because the primary observations are spectroscopic in nature,
dithering is mostly absent in our pure-parallel WFC3 data.  As a
result, detector artifacts (e.g., hot/warm pixels), uncorrected
cosmic rays, and WFC3/IR detector persistence must be carefully
considered in the data reduction and data analysis.  To mitigate
detector hot/warm pixels and cosmic rays, we employed a robust
algorithm based on a variation of Laplacian edge detection
\citep{vanDokkum2001} to filter the individual FLT files prior to
combining them with {\tt MultiDrizzle} \citep{Koekemoer2002}.  The
parameters for the algorithm were chosen to remove most sharp-edged
artifacts from the data, but to also be rather conservative so as not
to remove any real sources.

As detailed in \cite{Trenti2011}, our observing strategy is also
unique because it is designed to minimize the impact of WFC3/IR
detector persistence from observations taken in orbits immediately
prior to our exposures.  We take advantage of the property that the
amount of residual image persistence appears to decay roughly as a
power law with time.  Briefly, observations in either the \Jband\ band
or \Hband\ band are preceded in the same orbit by a comparably long
\Yband\ band exposure, and whenever possible, \Jband-band exposures
precede those in \Hband\ band\footnote{This Phase II strategy is
implemented in GO 11700 and GO 12752 (PI: Trenti), as well as the more
recent COS-GTO parallel observations.}.  Therefore any persistent
residual images from prior observations will be brighter in the bluer
bands and thus will not contaminate our high-redshift dropout sample.

{\tt MultiDrizzle} \citep{Koekemoer2002} was used to combine
individual exposures in a given filter to produce the final science
images as well as the associated inverse-variance weight maps.  The
images are drizzled to a final pixel scale of 0.08\arcsec/pixel.
Because BoRG is a pure-parallel program, the image depths vary among
our fields.  For a typical 4-orbit parallel field, we obtain exposures
of 2200~s, 3800~s, 1800~s, and 1800~s in \Vband, \Yband, \Jband, and
\Hband, respectively.  In a $0.4\arcsec$ diameter aperture, the
corresponding $5\sigma$ limiting magnitudes are approximately 27.4,
27.1, 26.9, and 26.6, respectively.  In Table~\ref{tbl:borg09}, we
list the survey fields, exposure times, and $5\sigma$ limiting
magnitudes ($r=0.32\arcsec$ aperture) of the initial BoRG09 dataset
\citep{Trenti2011}, and in Table~\ref{tbl:borg12} we list the same
properties for the new fields in the significantly expanded BoRG12
dataset.  We have released the reduced drizzled science and RMS images
for the BoRG dataset, which is publicly available via the BoRG website
at \href{http://wolf359.colorado.edu/}{wolf359.colorado.edu} and the
High Level Science Products on the MAST
website\footnote{\href{http://archive.stsci.edu/prepds/borg/}{archive.stsci.edu/prepds/borg}}

We used {\tt SExtractor} \citep{Bertin1996} in dual-image mode for
object detection and photometry.  For the detection image we used the
WFC3/IR \Jband\ data.  The inverse-variance weight images produced by
{\tt MultiDrizzle} were used to generate RMS maps for each image.  We
subsequently normalized the RMS maps to account for correlated noise
introduced by the drizzling procedure \citep[see][]{Trenti2011} and
used them in {\tt SExtractor} for both source detection and
photometry.  Sources were required to have at least 9 contiguous
pixels, each detected at a threshold of 0.7$\sigma$ above the
background.  Object colors and signal-to-noise values were measured
using isophotal magnitudes (ISOMAG) and total magnitudes were measured
within scalable Kron apertures (AUTOMAG).

The photometry in each field has been corrected for Galactic
extinction using the \cite{Schlegel1998} dust maps.  All measurements
given in this paper include this correction, which is typically modest
$A_{V}\lesssim 0.2$ because the survey primarily targets lines of
sight at high galactic latitudes.

\section{Selection of \lowercase{$z\sim8$} \Yband-band Dropout
Candidates}\label{sect:select}

We searched for $z\ga7.5$ galaxies using a \Yband-dropout selection
criteria in two broadband colors.  The general criteria are a strong
break between the \Yband\ and \Jband\ filters, a relatively blue or
flat \JmH\ color, and a non-detection in the optical \Vband\ (or
\Vbandb) band.  Specifically, following \cite{Trenti2011, Trenti2012a}
we require:

\begin{eqnarray*}
S/N_{Vband}       & < & 1.5  \\
(Y_{098}-J_{125}) & > & 1.75 \\
(J_{125}-H_{160}) & < & 0.02 + 0.15 [(Y_{098}-J_{125}) - 1.75]. \\
\end{eqnarray*}

These criteria select galaxies with redshifts in the range
$z\sim7.4-8.8$ (see Section~\ref{sect:uvlf}).  To minimize the
probability of contamination of our sample by low-redshift
interlopers, we impose a conservative non-detection threshold of
$1.5\sigma$ on the optical-band data.  The non-detection in the
optical $V$ band is fundamental to produce a clean sample of $z\sim8$
candidates (see \citealt{Bouwens2011b}).  If a source is not detected
in a given filter, we set its magnitude to the corresponding $1\sigma$
upper limit to calculate its colors.  For our final catalog, we
require sources to be detected with a signal-to-noise ($S/N$)
threshold of $S/N \geq 5$ in \Jband\ and $S/N \geq 2.5$ in \Hband, as
measured in isophotal apertures (ISOMAG).  For the bright $8\sigma$
catalog, we require sources to be detected with a signal-to-noise
threshold of $S/N \geq 8$ in \Jband\ and $S/N \geq 3.0$ in \Hband.

Using these color criteria, we identified 33 relatively bright
\Yband-dropout galaxy candidates with observed \Jband\ band magnitudes
between $25.5$ and $27.4$ mag over our 274~arcmin$^{2}$ search area.
Eight of these $z\sim8$ LBG candidates are brighter than $26.6$ mag
and are detected at a significance level of $>8\sigma$ in \Jband.
This is the largest sample of bright ($\Jband \la 27.4$) $z\sim8$
galaxies presented to date.  We list the properties of the candidates
in the $8\sigma$ and $5\sigma$ catalogs in Tables~\ref{tbl:8sig} and
\ref{tbl:5sig}, respectively.  The postage-stamp cutout images of the
eight $8\sigma$ candidates are shown in Figures~\ref{fig:stamps1} and
\ref{fig:stamps2}.

The (\YmJ) and (\JmH) colors of our $z\sim8$ candidates along with the
color-color selection criteria are illustrated in
Figure~\ref{fig:colcol}.  We also show the expected colors of galaxies
simulated over a wide range of redshifts and also those of low-mass
dwarf stars (e.g., \citealt{Knapp2004, Ryan2011}; Holwerda et al. 2012
(in prep)).  As Figure~\ref{fig:colcol} shows, our conservative $\YmJ
> 1.75$ color criterion is effective in minimizing contamination from
cool dwarf stars and low-redshift interlopers.

\subsection{Comparison with the Earlier BoRG09
Sample}\label{sect:borg09}

In this paper we employ an improved data reduction that includes
Laplacian filtering (see Section~\ref{sect:obs}).  Therefore, we
expect some photometric scatter within the measurement uncertainty
with respect to the previous catalogs of \Yband\ dropouts published in
\cite{Trenti2011, Trenti2012a}.  Indeed, this is the case:  one of the
four bright dropouts identified in \cite{Trenti2011}, source
``BoRG1k'', is now just marginally out of the catalog with
$Y_{098}-J_{125} = 1.70$.  This is not surprising, as that candidate
was at the edge of the selection window in the previous photometry and
we were already considering it likely ($p\sim 60\%$) to be a
contaminant that scattered into the dropout selection (see Section~5.2
in \citealt{Trenti2011}).  While it was, and still is, uncertain
whether this source is at $z\sim8$, our artificial source recovery
simulations (see Section~\ref{sect:uvlf}) take photometric scatter
into account when determining the effective volume of the survey,
making the derivation of the luminosity function robust.

With the new data reduction we also verified the photometry for field
BoRG58 (here borg\_$1437+5043$), where we identified a $z\sim8$
protocluster candidate \cite{Trenti2012a}.  The improved photometry
confirms the previous measurements, although we note that two of the
fainter $z\sim8$ candidates have scattered out of our current catalog.
In this new reduction and catalog, they are detected in the \Jband\
band at $4.98\sigma$ and $4.88\sigma$, slightly below our formal
threshold of $5\sigma$ detections (see Table~\ref{tbl:borg58}).  It is
worth noting that two new \Yband-dropout sources are also detected in
this field, just below the detection threshold, at $S/N \sim 4.5$ in
\Jband, providing circumstantial evidence that the overdensity extends
to fainter luminosities in line with our theoretical and numerical
predictions \citep{Trenti2012a}.  Deeper \HST\ imaging would be very
useful to further investigate the nature of this overdensity.

\subsection{Possible \lowercase{$z\sim8$} Candidates Excluded from the
Strict Sample}\label{sect:other}

While two additional candidates pass our $8\sigma$ selection criteria
(see Table~\ref{tbl:other}), we exclude them from our final catalog.
One of these sources (borg\_0240$-$1857\_392) is extremely compact and
clearly unresolved.  Given its compactness, the likely explanation is
that source is a Galactic star with either a significant fluctuation
in the photometric measurement or with an unusually red
$Y_{098}-J_{125}$ intrinsic color (at the $\sim 2\sigma$ level).
Because this object is located right above the Galactic Center ($l =
354.36830034\dg$, $b = 23.48948421\dg$), the source is most likely a
reddened L or T dwarf star.  An exciting, but much less likely,
alternative would be a $z\sim8$ QSO, although the limited area of BoRG
implies that this occurs with $p\lesssim 5\%$ based on the
\cite{Willott2010} QSO luminosity function predictions.

The other source we excluded (borg\_1632$+$3737\_386) is extremely
bright at 25.1 AB mag in the \Jband\ band and located immediately
adjacent ({\tt SExtractor} extraction flag ``2'') to an even brighter
($J_{125}=21.4$ mag) foreground spiral galaxy.  The spiral galaxy is
tidally interacting with a second spiral and has colors similar to the
dropout candidate ($Y_{098}-J_{125} = 1.2$, $J_{125}-H_{160}=0.35$,
$V_{606}-J_{125}=3.5$), but it is clearly detected in \Vband.  In
principle, the \Yband-dropout could have been lensed \cite[e.g.,
see][]{Wyithe2011Nat}, but because of the similar colors with the
foreground source, we consider it much more likely that this dropout
is part of the foreground system, with its \Jband-band photometry
partially contaminated by emission-line flux \citep[see][]{Atek2011}
and a \Vband\ continuum that cannot be detected in the current data
assuming its $V_{606}-J_{125}$ is similar to that of the spiral
galaxy.

\ifemulateapj\fige\fi
\ifemulateapj\figf\fi
\ifemulateapj\figg\fi
\ifemulateapj\tabg\fi

\section{The \lowercase{$z\sim8$} LBG Luminosity
Function}\label{sect:uvlf}

We derive the completeness, $C(m)$, and magnitude-dependent redshift
selection function, $S(z,m)$, of our dataset from simulations as
described in \cite{Oesch2009, Oesch2012}.  Briefly, artificial
galaxies with a range of spectral energy distributions (SEDs),
luminosities, redshifts, and sizes are added to the real images.  We
then rerun our detection and selection procedure on the data in each
individual field to determine both $C(m)$ and $S(z,m)$ (see
Figure~\ref{fig:szm}).  The simulations are based on $z\sim4$ galaxy
images rescaled to the desired input magnitude and to higher redshift
using standard evolutionary relations.  In particular we use a size
scaling of $(1+z)^{-1}$ as determined from LBGs in the range $z\sim
3-7$ \citep{Ferguson2004, Bouwens2004b, Oesch2010} and adopt a
UV-continuum slope distribution of $\beta = -2.5 \pm 0.4$, motivated
by the recent measurements at $z>6$ \cite{Bouwens2012,
Finkelstein2011, Dunlop2012}.  From these simulations we derive the
effective volume for our $z\sim8$ selection as a function of \Jband\
magnitude as shown in Figure~\ref{fig:effvol}.  By design these
simulations take into account the reduction of the effective area and
volume probed by the survey because of photometric scatter and
presence of foreground sources, including any persistence images in
the \Yband\ data.  We also note that because we are selecting sources
from the \Jband\ images, the $J_{125}-H_{160}$ color is not an
unbiased measure of the rest-frame UV slope $\beta$, so the
simulations we perform are crucial to derive the effective volume of
our selection, and show that we retain high efficiency, with
$S(z,m)\gtrsim0.8$ (Figure~\ref{fig:effvol}).

\ifemulateapj\tabh\fi
\ifemulateapj\tabi\fi

We construct the stepwise luminosity function from our BoRG
observations in $0.5$ mag bins for the sample of all $33$
\Yband-dropouts and for the smaller, but more robust, sample of $8$
sources with high $S/N$ ($\Jband\ >8\sigma$) detections.  The results
are shown in Table~\ref{tbl:swlf} and Figure~\ref{fig:uvlf} and take
into account $34\%$ contamination for the $S/N>8$ sample and $42\%$
contamination for the $S/N>5$ sample based on improved estimates
following \cite{Trenti2011, Trenti2012a}.  We take into account the
contamination levels of the legacy fields following \cite{Oesch2012}
and estimate the BoRG contamination rate by using the Early Release
Science (ERS) WFC3/IR data \citep{Windhorst2011}, which also used the
F098M filter as the $Y$ band, in combination with the GOODS F606W data
degraded to match our relative \Vband\ and \Yband\ depths.  The main
limitation of this approach is the limited area of the ERS data, which
is a factor $10$ smaller than the BoRG area.  Also, we assume that
contamination does not vary as a function of luminosity for a given
dataset, which is the working assumption of previous studies deriving
the LF at high redshift, both in legacy and pure-parallel fields.  In
principle, the contamination might vary with source brightness,
depending on the LF difference for contaminants and $z\sim8$ galaxies
with the same near-IR colors.  Without a spectroscopic follow-up
survey, and given the lack of confirmed $z\sim8$ galaxies even in
legacy surveys \citep{Schenker2012, Treu2012}, it is challenging, if
at all possible, to investigate this in more detail.  Yet, it is
reassuring that the spectra of the two BoRG $z\sim8$ candidates that
have been observed at Keck with NIRCAM by \cite{Schenker2012} and
\cite{Treu2012} allow us to exclude the scenario where these sources
are $z\sim1.5$ galaxies with strong emission lines, which are expected
to be the main contaminants of our selection (see \citealt{Atek2011}
and \citealt{Trenti2011}).

To explore the effect of contamination in BoRG, we performed the
experiment of leaving the contamination fraction as a free parameter
in the LF fit, as discussed below (see Table~\ref{tbl:contam}).  A
different contamination fraction would, under our assumptions, simply
scale up or down the BoRG LF measurement.  Shifting up our data by
$\sim 0.2$ dex, assuming no contamination, can be interpreted very
conservatively as an upper limit to the bright end of the $z\sim8$
galaxy luminosity function.  As Figure~\ref{fig:uvlf} shows, the low-
and high-$S/N$ samples provide a consistent determination of the
luminosity function of \Yband\ dropouts in the BoRG dataset and the
data in our faintest bins agree with (and are actually slightly lower
than) the brightest bins in \cite{Bouwens2011b} from ERS and HUDF09
data.  Finally, the measure of a strong clustering signal from the
BoRG survey argues against a strong contamination fraction
\cite{Trenti2012a}.

We derive Schechter function parameters by performing a
maximum-likelihood fit to the data, assuming a Poisson distribution of
the galaxy number counts in each magnitude bin.  Specifically, we
maximize the Poisson likelihood for observing $N^{obs}$ sources in a
given magnitude bin when $N^{exp}$ are expected to be observed from a
given Schechter LF.  The likelihood $\mathcal{L}$ is expressed as
$\mathcal{L} = \Pi_{j} \Pi_{i}\ P(N^{obs}_{j,i}, N^{exp}_{j,i})$,
where $P(N^{obs}, N^{exp})$ is the Poisson probability distribution
and the products are taken over all the fields, $j$, and magnitude
bins, $i$.  To establish the best estimate of the overall shape of the
UV LF, we combine our BoRG dataset with those from the deeper
ERS+HUDF09 data of \cite{Bouwens2011b}, providing the largest dynamic
range in luminosity currently possible.  Given that the CANDELS
observations yield very different preliminary results depending on the
team that analyzed the data to search for $z\sim8$ galaxies
\citep[see][]{Yan2012, Oesch2012}, we exclude those data from our
analysis.

\ifemulateapj\figh\fi
\ifemulateapj\figi\fi

At $68\%$ confidence, for $h=0.7$ we derive $\phi_{*} =
(4.3^{+3.5}_{-2.1}) \times 10^{-4}$ Mpc$^{-3}$, $\Mstar =
-20.26^{+0.29}_{-0.34}$, and a very steep faint-end slope $\alpha =
-1.98^{+0.23}_{-0.22}$.  As observed in Figure~\ref{fig:uvlf}, overall
the best fit provides a very good description of the data.  The
covariance in the parameter values is smaller than that in the
CANDELS+ERS+HUDF analysis by \cite{Oesch2012} (\cite{Yan2012} do not
fit a LF to their data), but it is still significant, as shown in
Figure~\ref{fig:lfparam}.  From this figure it is also clearly evident
that our \Mstar\ versus $\alpha$ parameter values have much better
constraints than previous studies at $z\sim8$ \citep{Oesch2012}.
Formally the bright end of the luminosity function has a marginally
low $\Mstar$ value.  However, this is compensated by the low
$\phi_{*}$ such that the best-fit Schechter function ($\phi_{*} =
5.9^{+10.1}_{-3.7} \times 10^{-4}$~Mpc$^{-3}, M_{*} = -20.1 \pm 0.52$)
derived by \cite{Bouwens2011b} is fully consistent with our
determination at $68\%$ confidence.  As can be seen in
Figure~\ref{fig:uvlf}, the faintest three bins of the BoRG dataset
match very closely the brightest bins in \cite{Bouwens2011b}.  The
latest BoRG LF is also very similar to our earlier fit from
\cite{Trenti2011}, where we had kept $\phi_{*}$ and $\alpha$ fixed
(see Table~\ref{tbl:lfparam}).

Table~\ref{tbl:contam} shows the impact of varying the contamination
fraction $f$ for the BoRG dataset and its associated contamination
correction $1/(1-f)$ that is applied to the intrinsic counts per
magnitude bin to obtain the estimated observed counts.  The table
shows that the impact of contamination around our fiducial value
$f=0.42$ is relatively modest.  We note that the highest likelihood is
associated to a contamination fraction just below our fiducial value,
providing indirect evidence to suggest that our derived contamination
fraction is robust.  Varying the contamination fraction around
$f=0.33$ has almost no impact on the determination of $M_{*}$, as the
LF is well sampled by the BoRG data in this luminosity range, but
affects mostly $\alpha$; if we have underestimated $f$, then the
faint-end slope would be even steeper.

As shown in Table~\ref{tbl:lfparam}, our determination of the $z\sim8$
luminosity function is consistent with previous work
\citep{McLure2010,Lorenzoni2011,Trenti2011,Bouwens2011b,Oesch2012},
taking into account the significant uncertainties in all these
measures.  Formally, our best fit prefers a low normalization for
$\phi_{*}$ and a brighter $M_{*}$ (similar to the $z\sim7$ value), but
the covariance between these parameters is very large (see
Figure~\ref{tbl:lfparam}).  In particular our BoRG+ERS+HUDF LF fit
provides a better constraint on the \Mstar vs. $\alpha$ parameter
uncertainties than the CANDELS+ERS+HUDF determination at $z\sim8$
\cite{Oesch2012}.

In comparing different datasets, it is important to note that the
bright end is still sparsely sampled; for example, the brightest BoRG
magnitude bin includes only one source.  Therefore, small-number
fluctuations are very significant and Poisson noise in the brightest
bins can have a large impact on the fit.  However, our maximum
likelihood fit shows, independent of the $M_{*}$ versus $\phi_{*}$
degeneracy, that the faint-end slope $\alpha$ is being constrained
with growing accuracy.  Of note, we reduce its $1\sigma$ uncertainty
at $z\sim8$ to $\pm 0.2$ and show in Figure~\ref{fig:alpha} that the
steepening trend suggested by previous studies
\citep[e.g.,][]{Bouwens2011b} is also supported by our LF fit.

The steep faint-end slope derived in our best-fit LF implies that
faint galaxies are playing a key role as major producers of ionizing
photons.  For a Schechter LF, most of the luminosity density
contribution is at the faint end (e.g., see left panel of Fig. 3 in
\citealt{Trenti2010}).  Furthermore, in the case of $\alpha \sim -2$,
there is a logarithmically divergent contribution from sources below
the detection limit of the survey.  This implies that the total
ionizing flux produced by galaxies is sensitive to the exact measure
of $\alpha$ (as well as on the extrapolation beyond the detection
limit), as shown, for example, in the left panel of Fig. 4 of
\citet{Bouwens2012}.  Assuming that star formation continues to be
efficient in lower luminosity galaxies down to $M_{AB}\sim -12$ at
$z\sim 8$ (the limit of atomic hydrogen cooling halos, e.g. see
\citealt{Trenti2010, Finlator2011}), then we expect that the current
\HST\ observations are only observing $\sim 20\%$ of the light present
at $z \sim 8$.  A direct proof that we are detecting only the tip of
the iceberg of star formation during the epoch of reionization is
provided by \HST\ observations of high-redshift, spectroscopically
confirmed GRBs that have failed to detect host galaxies despite
reaching ultrafaint sensitivity ($M_{AB}\sim -17$, see
\citealt{Tanvir2012}).  The non-detection of galaxies at locations in
the sky where it is confirmed that star formation is happening
(because a GRB explosion has occurred) implies that $M_{AB}\gtrsim
-15$ galaxies were indeed the main ionizing sources
\citep{Trenti2012b}, in full agreement with the interpretation of the
LF fit derived in this paper.

Interestingly, our determination of the $z\sim8$ luminosity function
at the bright end is between the debated measurements from the CANDELS
dataset obtained by \cite{Oesch2012} and \cite{Yan2012}.  However, our
data clearly show a well-behaved Schechter function, similar to
\cite{Oesch2012}, without the unusual shape derived by \cite{Yan2012}.
Our LF determination and that by \cite{Oesch2012} are consistent at
the $\sim 2\sigma$ level.  The observed differences could be partially
related to the environment (i.e. cosmic variance \citep{Trenti2008}).
\cite{Oesch2012} discuss that the ERS measurement by
\cite{Bouwens2011b} could have been affected by an overdensity.
However, given that the BoRG dataset is not affected by cosmic
variance and agrees with \cite{Bouwens2011b}, it may well be that the
CANDELS-South field is underdense, which would explain the lower
$M_{*}$ value derived from that dataset.  The upcoming BoRG
observations and those in the CANDELS-North field will help clarify
the nature of the discrepancy and provide further improvements to the
determination of the bright end of the $z\sim8$ luminosity function.

\section{Summary and Conclusions}\label{sect:concl}

We present the discovery of 33 Lyman break galaxy (LBG) candidates at
$z\sim8$ detected in \HST/WFC3 imaging as part of the Brightest of
Reionizing Galaxies (BoRG) pure-parallel survey.  Our sample of bright
\Yband-dropout galaxy candidates have \Jband-band magnitudes between
$25.5$ and $27.4$ mag, obtained from $59$ independent lines of sight
over a total area of $274$ arcmin$^2$.  The pure-parallel nature of
BoRG allows us to obtain an estimate of the galaxy luminosity function
which is unaffected by large-scale structure uncertainty and distinct
from determinations using legacy surveys limited to a single, or a
few, contiguous fields.

With our new data we detect galaxies between $-22\lesssim
M_{AB}\lesssim -19.75$, demonstrating the bright end of the $z\sim8$
luminosity function is well described by a Schechter form, similar to
that found at lower redshifts.  Our measurement of the number density
of galaxies near the BoRG detection limit ($M_{AB}\sim -20$) is in
agreement with the results by \cite{Bouwens2011b}.  Their measurement
is based on ERS+HUDF09 observations that extend to much fainter
luminosities, but they lack the area to detect the brighter galaxies
at $M_{AB} \lesssim -20.5$ we identify in the BoRG survey.  The two
datasets are complementary and allow us to obtain the best
determination yet of the galaxy luminosity function at $z\sim8$, with
$\phi_{*} = (4.3^{+3.5}_{-2.1}) \times10^{-4}$ Mpc$^{-3}$, $\Mstar =
-20.26^{+0.29}_{-0.34}$, and $\alpha = -1.98^{+0.23}_{-0.22}$ for
$h=0.7$.

Covariance in the parameters, albeit reduced thanks to the increase in
the dynamic range of the fit, is still very significant.  However,
this affects primarily \Mstar\ versus $\phi_{*}$.  We find that our
BoRG+ERS+HUDF LF fit provides a better constraint on the \Mstar\ versus
$\alpha$ parameter uncertainties than the CANDELS+ERS+HUDF
determination at $z\sim8$ \cite{Oesch2012}.  The faint-end slope
uncertainty is starting to be reduced to a point where there is a hint
of steepening compared to $z\la6$.  This steepening is expected based
on numerical and theoretical models of galaxy formation in the epoch
of reionization \citep{Trenti2010, Jaacks2012}.  As a consequence of
this large abundance of faint galaxies, it is expected that such
systems will dominate the total star formation rate and ionizing
photon production \citep{Shull2012}, which has recently been confirmed
observationally by the non-detection of host galaxies in a sample of
six $z>5$ Gamma-Ray Bursts \citep{Trenti2012b, Tanvir2012}.  Our
determination of a steep $\alpha$ for a Schechter fit of the LF is
robust against our estimate of the contamination fraction of the BoRG
survey ($f=0.42$ for the $5\sigma$ sample); as shown in
Table~\ref{tbl:contam}, if $f$ were higher, $\alpha$ would be steeper.
In the near future, the upcoming ultradeep observations of GO 12498
(PI Ellis) will improve the determination of the number density of the
faintest galaxies observable by \HST, while scheduled BoRG and
CANDELS-North observations will further increase the search area for
the brightest galaxies at $z\sim8$, allowing us to further tighten the
constrains on the LF.

The $z\sim8$ candidates identified here are good targets for follow-up
observations from ground and space observatories.  We have started a
spectroscopic campaign to confirm their redshift, to measure the
distribution of Ly$\alpha$ equivalent width (which is related to the
IGM ionization state), and also to rule out contamination from
emission-line galaxies at $z\sim1.5$ \citep{Treu2012}.  The
availability of multi-object spectrographs both on Keck (MOSFIRE) and
Gemini South (FLAMINGOS-2), will enable faster progress, especially in
fields that have overdensities of sources similar to the protocluster
candidate we reported in \cite{Trenti2012a}.  In addition, the
brightest BoRG sources are expected to have star formation rates of
$10-20$~\Msunyr\ based on their rest-frame UV luminosity, making them
prime targets for ALMA observations as they are expected to fall well
within the telescope sensitivity based on the \cite{Carilli2008}
predictions.  If these candidates are spectroscopically confirmed,
this would allow us to possibly extend the detection of
\CII~158~$\mu$m, high-$J$ CO emission lines, and perhaps dust emission
as well, to $z\sim8$, after the recent record established at $z=7.1$
from a QSO host galaxy \citep{Venemans2012}.  Finally, \Spitzer/IRAC
observations, combined with spectroscopic redshifts, would also be
very useful to quantify the rest-frame optical properties of these
sources before the advent of the \textit{James Webb Space Telescope}.

\acknowledgments
The {Space Telescope Science Institute} is operated by AURA Inc.,
under NASA contract NAS5-26555.  This work was supported by grants
HST-GO-11563, HST-GO-11700, and HST-GO-12572.  This work included
coordinated parallel data taken as part of the COS GTO program (PI
Green).  Support for PO was provided by NASA through Hubble Fellowship
grant HF-51278.01.  JS acknowledges support from grant NNX08-AC14G.
MT gratefully acknowledges the hospitality of the Kavli Institute for
Theoretical Physics, where part of this research was carried out with
support by the National Science Foundation under Grant
No.~PHY11-25915.  Part of the work presented in this paper was
performed by MT and TT while attending the program ``First Galaxies
and Faint Dwarfs: Clues to the Small Scale Structure of Cold Dark
Matter" at the Kavli Institute of Theoretical Physics at the
University of California Santa Barbara.

%------------------------------------------------------------------------------
% Bibliography
%------------------------------------------------------------------------------
\bibliographystyle{apj}
%\bibliography{highz}
%\input{ms.bbl}

%------------------------------------------------------------------------------
% Tables and Figures
%------------------------------------------------------------------------------
\ifemulateapj\else
    \taba
    \clearpage
    \tabb
    \clearpage
    \tabc
    \clearpage
    \tabd
    \clearpage
    \tabe
    \clearpage
    \tabf
    \clearpage
    \tabg
    \clearpage
    \tabh
    \clearpage
    \tabi
    \clearpage

    \figa
    \clearpage
    \figb
    \clearpage
    \figc
    \clearpage
    \figd
    \clearpage
    \fige
    \clearpage
    \figf
    \clearpage
    \figg
    \clearpage
    \figh
    \clearpage
    \figi
\fi

\end{document}